\documentclass[aps,prd,floats,superscriptaddress,floatfix,nofootinbib,preprintnumbers,notitlepage,eqsecnum]{revtex4-1}
\usepackage{amsmath}
\usepackage{amsfonts}
\usepackage{graphicx}
\usepackage{slashed}
\usepackage{dcolumn}
\usepackage{bm}
\usepackage{amssymb}
\usepackage{latexsym}
\usepackage{color}
\usepackage{url}

\newcommand{\pbrac}[1]{\left( #1 \right)}
\newcommand{\tbrac}[1]{\left[ #1 \right]}
\newcommand{\cbrac}[1]{\left\{ #1 \right\}}

\begin{document}

\title{Vacuum Stability and Triviality Analyses of the Renormalizable Coloron Model}

\author{R. Sekhar Chivukula}
\email[]{sekhar@msu.edu}
\affiliation{Department of Physics and Astronomy, Michigan State University, East Lansing, Michigan 48824, USA}
\author{Arsham Farzinnia}
\email[]{farzinnia@ibs.re.kr}
\affiliation{Center for Theoretical Physics of the Universe\\Institute for Basic Science (IBS), Daejeon 305-811, Republic of Korea}
\author{Elizabeth H. Simmons}
\email[]{esimmons@msu.edu}
\affiliation{Department of Physics and Astronomy, Michigan State University, East Lansing, Michigan 48824, USA}

\preprint{CTPU-15-02}
\preprint{MSUHEP-150412}

\date{\today}

\begin{abstract}
The renormalizable coloron model is built around a minimally extended color gauge group, which is spontaneously broken to QCD. The formalism introduces massive color-octet vector bosons (colorons), as well as several new scalars and fermions associated with the symmetry breaking sector. In this paper, we examine vacuum stability and triviality conditions within the context of the renormalizable coloron model up to a cutoff energy scale of 100~TeV, by computing the $\beta$-functions of all relevant couplings and determining their running behavior as a function of the renormalization scale. We constrain the parameter space of the theory for four separate scenarios based on differing fermionic content, and demonstrate that the vectorial scenarios are less constrained by vacuum stability and triviality bounds than the chiral scenarios.  Our results are summarized in exclusion plots for the separate scenarios, with previous bounds on the model overlaid for comparison.  We find that a 100~TeV hadron collider could explore the entire allowed parameter space of the chiral models very effectively.  
\end{abstract}

\maketitle

\section{Introduction}\label{Intro}

As we anticipate the flood of new data from the second run of the CERN Large Hadron Collider, thoughts are turning to the new states that may be discovered, states that would represent a first clear sign of physics beyond the Standard Model (SM).  Many theories of new physics that attempt to address unresolved challenges, including flavor physics and naturalness, incorporate an extended strong interaction sector.  
Examples include models such as topcolor \cite{Hill:1991at}, the flavor-universal coloron \cite{Chivukula:1996yr}, chiral color \cite{Frampton:1987dn,Bagger:1987fz}, chiral color with unequal gauge couplings \cite{Martynov:2009en}, flavor non-universal chiral color \cite{Frampton:2009rk} and a flavorful top-coloron model \cite{Top-Coloron}. In such models, the color gauge group is extended to $SU(3)_{1\, c} \times SU(3)_{2\, c}$ at high energies; this enlarged gauge group is spontaneously broken at low energies to the diagonal $SU(3)_{c}$ subgroup, which is identified with QCD.   In the process of spontaneous symmetry breaking, the broken generators become identified with a set of massive color-octet gauge bosons, generically called \textit{colorons}.  

As discussed in \cite{recom,Chivukula:1996yr,Bai:2010dj,Chivukula:2013xka}, one can construct a renormalizable theory by including scalars transforming appropriately under the extended color symmetry.  In addition to the new massive color gauge states, new colored and uncolored scalar degrees of freedom are present in this model. Moreover, cancelation of potential anomalies, which would arise if the couplings of the ordinary quarks to colorons were chiral, can require the existence of new spectator fermions. Hence, if this formalism corresponds to the correct description of nature, a rich spectrum of new scalar, fermionic, and vector particles with novel properties are predicted, the discovery of which may lie within the reach of the LHC.

A first complete study of hadron collider production of colorons at next-to-leading order was presented in \cite{Chivukula:2011ng,Chivukula:2013xla}. The scalar sector of the renormalizable coloron model has been studied in \cite{Chivukula:2013xka}, and bounds on its properties were set by imposing limits arising from the global minimum of the potential coinciding with the scalar VEVs, unitarity, electroweak precision tests, and properties of the 125~GeV Higgs-like scalar discovered in 2012 at the LHC \cite{Aad:2012tfa,Chatrchyan:2012ufa}. A subsequent paper \cite{Chivukula:2014rka} examined the properties of the additional (heavier) color-singlet  CP-even scalar boson in the model, and established the expected reach of the $\sqrt s = 14$~TeV LHC for this state in three selected scenarios with 0, 1, and 3 spectator fermion generations, respectively.  

In this paper, recalling that gauge theories with extended scalar sectors are often subject to Landau poles below the Planck scale \cite{GenBeta1,GenBeta2}, we examine constraints that considerations of vacuum stability and triviality place on the renormalizable coloron model.  After introducing the essential features of the model in Section~\ref{Prelim}, we establish the separate cases within the model that we propose to examine in Section~\ref{Vacstab}.  Specifically, we lay out four distinct scenarios with differing numbers of spectator fermion generations and different origins of the spectator fermions' masses.  We then compute the $\beta$-functions of all relevant couplings and determine their running behavior as a function of the renormalization scale up to a cutoff energy scale of order 100 TeV.  Details regarding the form of the $\beta$-functions in the different scenarios are summarized in a set of three Appendices.  The results of the calculations and their implications for the parameter space of the model are discussed in Section~\ref{Resl}.  We show how the behavior of the $\beta$-functions translates into constraints on the model as a function of the physically relevant free parameters: masses, scalar vacuum expectation value (VEV), and scalar mixing angle.  We also provide exclusion plots showing how the constraints arising from vacuum stability and triviality compare with those obtained from other theoretical and phenomenological sources in prior work \cite{Chivukula:2013xka,Chivukula:2014rka}.  The concluding section summarizes our findings, most notably that the scenarios in which the spectators' couplings to colorons are vectorial are far less constrained  by the vacuum stability and triviality bounds than those in which the couplings are chiral.  The latter scenarios could be explored quite thoroughly at a 100 TeV hadron collider.

\section{Formalism}\label{Prelim}

In this section, we briefly discuss the renormalizable coloron model \cite{Chivukula:1996yr,recom,Bai:2010dj,Chivukula:2013xka,Chivukula:2014rka} and review the necessary notation and definitions that we will employ  \cite{Chivukula:2013xka,Chivukula:2014rka}. In the framework of the renormalizable coloron model, the color gauge group of the standard model (SM) is enlarged to $SU(3)_{1\, c} \times SU(3)_{2\, c}$; this extended group is spontaneously broken to its diagonal subgroup, $SU(3)_{c}$, which we identify with ordinary QCD.  The spontaneous symmetry breaking also produces a set of additional massive color-octet vector bosons, generically referred to as \textit{colorons} in this work. The colorons obtain their mass by ``eating'' the colored Nambu-Goldstone bosons of the eight broken generators of the original extended color gauge group. Hence, the full theory may be characterized schematically as
\begin{equation}\label{gaugestr}
SU(3)_{1\, c} \times SU(3)_{2\, c} \times SU(2)_{L} \times U(1)_{Y} \longrightarrow SU(3)_{c} \times SU(2)_{L} \times U(1)_{Y} \longrightarrow SU(3)_{c} \times U(1)_{\text{EM}} \ ,
\end{equation}
where the gauge symmetry breaking in the electroweak and the color sectors occurs at different energy scales.

\subsection{The Boson Sector}\label{Boso}

The spontaneous symmetry breaking in the (unaltered) SM electroweak sector, $SU(2)_{L} \times U(1)_{Y} \to U(1)_{\text{EM}}$, is facilitated by the usual (color-singlet) Higgs field doublet $\phi$, at the weak scale $v_{h} = 246$~GeV:
\begin{equation}\label{phi}
\phi= \frac{1}{\sqrt{2}}
\begin{pmatrix} \sqrt{2}\,\pi^+ \\ v_h+h_0+i\pi^0 \end{pmatrix} \ ,
\end{equation}
where $h_{0}$ corresponds to the SM Higgs boson, and $\pi^{0,\pm}$ are the electroweak Nambu-Goldstone bosons eaten by the $Z$ and the $W^{\pm}$ gauge fields. The extended strong sector, on the other hand, is spontaneously broken to the diagonal QCD subgroup, $SU(3)_{1\, c} \times SU(3)_{2\, c} \to SU(3)_{c}$, via the vacuum expectation value of a complex multicomponent (electroweak-singlet) scalar $\Phi$, which assumes a bi-fundamental form $(3,\bar{3})$ under the two original color groups \cite{Bai:2010dj,Chivukula:2013xka}:
\begin{equation}\label{Phi}
\Phi = \frac{1}{\sqrt{6}} \pbrac{v_{s} + s_{0} + i {\mathcal A}} {\cal I}_{3\times 3} + \pbrac{G^a_H + i G^a_G}t^a \qquad \pbrac{t^a \equiv \lambda^a/2} \ .
\end{equation}
Here, $s_{0}$ ($\mathcal A$) denotes a gauge-singlet scalar (pseudoscalar) degree of freedom, $G^a_H$ represents a set of massive color-octet scalars, and $\lambda^{a}$ is the set of Gell-Mann matrices. The energy scale at which the spontaneous symmetry breaking occurs in the extended color sector is characterized by $v_{s}$---the non-zero VEV of the gauge-singlet scalar $s_{0}$---which is presumed to be higher than the scale of electroweak symmetry breaking ($v_{s} > v_{h}=246$~GeV).

The color-octet pseudoscalars  $G^a_G$ in \eqref{Phi} represent the colored Nambu-Goldstone bosons, which are eaten by the colorons to induce their mass \cite{Bai:2010dj,Chivukula:2013xka}:
\begin{equation} \label{MC}
M_{C}^{2} = \frac{v_{s}^{2}}{6} \pbrac{g_{s_{1}}^{2} + g_{s_{2}}^{2}} \ ,
\end{equation}
with $g_{s_{i}}$ the corresponding couplings of the original $SU(3)_{i\,c}$ color gauge groups ($i=1,2$). Furthermore, the usual QCD coupling $g_{s}$ of the massless gluons may be expressed in terms of the $g_{s_{i}}$ couplings, according to \cite{Bai:2010dj,Chivukula:2013xka,Chivukula:2011ng}:
\begin{equation}\label{gs}
\frac{1}{g_s^2} = \frac{1}{g_{s_1}^2}+\frac{1}{g_{s_2}^2}\ .
\end{equation}
One notes that the relations \eqref{MC} and \eqref{gs} may be inverted to express the gauge couplings $g_{s_{1}}$ and $g_{s_{2}}$ in terms of the physically more relevant quantities $M_{C}$, $v_{s}$, and $g_{s}$, resulting in:
\begin{equation}\label{gs12Match}
g_{s_{1,2}}^{2} = \frac{3 M_{C}^{2}}{v_{s}^{2}} \tbrac{1 \mp \sqrt{1-\frac{2 g_{s}^{2} \, v_{s}^{2}}{3M_{C}^{2}}}} \ .
\end{equation}
As an immediate consequence of \eqref{gs12Match}, one deduces a theoretical lower bound for the coloron mass: \begin{equation}\label{MClow}
M_{C} \geq \sqrt{2/3}\, g_{s} v_{s}\ .
\end{equation}

The scalar sector of the model is described by the Lagrangian:
\begin{equation}\label{Lscal}
\mathcal{L}_{\text{scalar}} = D^\mu \phi^\dagger D_\mu \phi+{\rm Tr} \tbrac{D^\mu\Phi^{\dagger} D_\mu\Phi} - V(\phi,\Phi) \ ,
\end{equation}
with the usual electroweak covariant derivative acting on $\phi$ \eqref{phi}. Defining $G_{i\, \mu}^{a}$ as the gauge fields of the two original $SU(3)_{i\,c}$ color groups,\footnote{Note that the gluon and the coloron are the corresponding physical gauge boson definitions, once $G_{1\mu}^{a}$ and $G_{2\mu}^{a}$ are orthogonally rotated (with a mixing angle $\theta_{c}$) into their mass eigenstate basis (see e.g. \cite{Chivukula:2011ng,Chivukula:2013xla} for a detailed treatment). The theoretically ``lightest'' coloron mass, $M_{C} = \sqrt{2/3}\, g_{s} v_{s}$ (c.f. \eqref{MClow}), corresponds to the case of maximal mixing between the two gauge groups $SU(3)_{i\, c}$ ($\theta_{c} = \pi/4$), and equal original gauge couplings, $g_{s_{1}}=g_{s_{2}}$; the \textit{axigluon} \cite{Frampton:1987dn,Bagger:1987fz} is an example of this case.} the color covariant derivative acting on the bi-fundamental $\Phi$ \eqref{Phi} takes the form:
\begin{equation}\label{Colcoder}
D_\mu \Phi = \partial_\mu \Phi - i g_{s_1} G^a_{1\mu} t^a \Phi + i g_{s_2} \Phi\,  G^a_{2\mu} t^a \ .
\end{equation}
As discussed in \cite{Bai:2010dj,Chivukula:2013xka}, the most general renormalizable form of the scalar potential contains the following terms:
\begin{equation}\label{pot}
\begin{split}
V(\phi,\Phi) = &\, \frac{\lambda_h}{6}\pbrac{\phi^\dagger \phi - \frac{v^2_h}{2}}^2 + \lambda_m\pbrac{\phi^\dagger \phi - \frac{v^2_h}{2}} \pbrac{{\rm Tr}\tbrac{\Phi^\dagger \Phi} - \frac{v^2_s}{2}}\\
&+ \frac{\lambda_s}{6}\pbrac{{\rm Tr}\tbrac{\Phi^\dagger \Phi}}^2 + \frac{\kappa_s}{2} {\rm Tr}\tbrac{\pbrac{\Phi^\dagger \Phi}^{2}} -\frac{\lambda_s + \kappa_s}{\sqrt{6}}\, v_{s} r_\Delta  \pbrac{{\rm Det}\Phi + {\rm h.c.}}  -\frac{\lambda_s + \kappa_s}{6} \, v_s^{2} \pbrac{1 - r_{\Delta}} {\rm Tr}\tbrac{\Phi^\dagger \Phi} \ ,
\end{split}
\end{equation}
with the five real and dimensionless couplings $\lambda_{h}$, $\lambda_{m}$, $\lambda_{s}$, $\kappa_{s}$, and $r_{\Delta}$. The last of these is always accompanied by powers of the singlet scalar VEV, $v_{s}$, in the potential and, hence, does not participate in the quartic scalar interactions. The potential is bounded from below under the quartic coupling conditions \cite{Chivukula:2013xka}:
\begin{equation}\label{stab}
\lambda_h > 0 \ , \qquad \lambda_s^{\prime} > 0 \ , \qquad \kappa_{s} > 0 \ , \qquad \lambda_m^2 < \frac{1}{9} \lambda_h \lambda_s^{\prime} \ ,
\end{equation}
where we define $\lambda_{s}^{\prime} \equiv \lambda_{s} + \kappa_{s}$ for later convenience. In addition, demanding that the global minimum of the potential \eqref{pot} coincides with the scalar VEVs, $\langle \phi \rangle = \dfrac{v_h}{\sqrt{2}} \begin{pmatrix} 0 \\ 1\end{pmatrix}$ and $\langle \Phi \rangle = \dfrac{v_s}{\sqrt{6}} \,\mathcal I_{3\times 3}$ (c.f. \eqref{phi} and \eqref{Phi}), imposes a nontrivial condition on the $r_{\Delta}$ coupling \cite{Chivukula:2013xka}:
\begin{equation}\label{rdel}
0 \le r_{\Delta} \le \frac{3}{2} \ .
\end{equation}

From the scalar potential \eqref{pot}, one deduces the following expressions for the masses of the singlet pseudoscalar and the scalar color-octet \cite{Bai:2010dj,Chivukula:2013xka}:
\begin{equation}\label{mAmGH}
m_{\cal A}^{2} = \frac{v_{s}^{2} }{2} \, r_{\Delta} \lambda_{s}^{\prime} \ , \qquad m_{G_{H}}^{2} = \frac{1}{3} \pbrac{v_{s}^{2} \, \kappa_{s}+ 2 m_{\cal A}^{2}} \ ,
\end{equation}
which implies
\begin{equation}\label{mAcond}
m_{\mathcal A} \leq \sqrt{\frac{3}{2}} \, m_{G_{H}} \ .
\end{equation}

The two massive scalars in the potential that have non-zero VEVs, $h_{0}$ and $s_{0}$, are mixed with one another due to the mixing coupling $\lambda_{m}$, and may be diagonalized by performing an orthogonal rotation into their mass eigenstate basis:
\begin{equation}\label{massbasis}
\begin{pmatrix} h_0\\ s_0 \end{pmatrix}
=  \begin{pmatrix} \cos\chi & \sin\chi \\ -\sin\chi & \cos\chi \end{pmatrix} \begin{pmatrix} h \\ s \end{pmatrix} \ , \qquad \cot 2\chi \equiv \dfrac{1}{6\lambda_m} \tbrac{ \lambda_s^{\prime}\pbrac{1-\dfrac{r_{\Delta}}{2}} \dfrac{v_s}{v_h} - \lambda_h \dfrac{v_h}{v_s} } \ .
\end{equation}
Consequently, the mass eigenstates $h$ and $s$ constitute the corresponding physical scalars of the theory with the masses:
\begin{equation}\label{mhs}
m_{h,s}^{2} = \frac{1}{6} \cbrac{ \lambda_h v_h^2 + \lambda_s^{\prime} v_s^2\pbrac{1-\frac{r_{\Delta}}{2}} \pm \tbrac{\lambda_h v_h^2 - \lambda_s^{\prime} v_s^2\pbrac{1-\frac{r_{\Delta}}{2}}}\sec 2\chi } \ .
\end{equation}
The relations \eqref{mAmGH}, \eqref{massbasis}, and \eqref{mhs} may be used to convert the five dimensionless couplings of the potential \eqref{pot} into the physically more relevant quantities $\sin\chi$, $v_{s}$, $m_{h}$, $m_{s}$, $m_{\mathcal A}$, and $m_{G_{H}}$, according to:
\begin{equation}\label{scalarMatch}
\begin{split}
\lambda_{h} = &\, \frac{3}{2}\frac{m_{h}^{2}+m_{s}^{2}+\pbrac{m_{h}^{2}-m_{s}^{2}}\cos 2\chi}{v_{h}^{2}} \ , \quad\qquad\qquad\qquad\,\,\lambda_{m} = \frac{1}{2}\frac{m_{s}^{2}-m_{h}^{2}}{v_{h} v_{s}}\sin 2\chi \ , \\
\lambda_{s}^{\prime} = &\, \frac{1}{2}\frac{2m_{\mathcal A}^{2}+3\pbrac{m_{h}^{2}+m_{s}^{2}}-3\pbrac{m_{h}^{2}-m_{s}^{2}}\cos 2\chi}{v_{s}^{2}} \ , \qquad \kappa_{s} = \frac{3m_{G_{H}}^{2}-2m_{\mathcal A}^{2}}{v_{s}^{2}} \ , \\
r_{\Delta} = &\, \frac{4 m_{\mathcal A}^{2}}{2m_{\mathcal A}^{2}+3\pbrac{m_{h}^{2}+m_{s}^{2}}-3\pbrac{m_{h}^{2}-m_{s}^{2}}\cos 2\chi} \ .
\end{split}
\end{equation}
Note that an attractive or repulsive interaction between the two scalar fields $\phi$ and $\Phi$ in the potential \eqref{pot}, characterized by the sign of $\lambda_{m}$, is reflected in the sign of the mixing angle $\sin\chi$. Moreover, as evident from the potential \eqref{pot}, the two scalar fields decouple in the limit $\lambda_{m} \to 0$. Based on \eqref{rdel} and the expression for $r_{\Delta}$ in \eqref{scalarMatch}, one also obtains the condition
\begin{equation}\label{rdelthresh}
m_{\mathcal A}^2 \leq 9\tbrac{m_h^2 \sin^2 \chi + m_s^2 \cos^2\chi}  \ ,
\end{equation}
which results in the lower bound: $m_s \ge \frac{1}{3}\, m_{\mathcal A}$, in the decoupling limit $(\lambda_{m} \to 0$).

In this framework, the lighter of the two color-neutral physical scalars in \eqref{massbasis} is assumed to be the $h$ degree of freedom, which is identified with the Higgs-like state discovered at the LHC \cite{Aad:2012tfa,Chatrchyan:2012ufa}; i.e., $m_{h}=125$~GeV. In addition, the theoretical and experimental analyses of \cite{Chivukula:2013xka,Chivukula:2014rka} disfavor large values of the mixing angle $\sin\chi$, implying that the $h$~scalar is more ``SM-like'', while the heavier $s$~boson appears more ``singlet-like''. The same analyses prefer a singlet VEV $v_{s} \gtrsim 1$~TeV. Tevatron searches have already excluded scalar color-octet bosons in the mass range $50 \lesssim m_{G_{H}} \lesssim 125$~GeV \cite{Aaltonen:2013hya} and LHC searches place the lower bound on the masses of scalar color-octet bosons at 2.70 TeV  \cite{Aad:2014aqa,Khachatryan:2015sja}; we, therefore, assume $m_{G_{H}} > m_{h}$ throughout this work, in accordance with the assumptions made in the previous studies \cite{Chivukula:2013xka,Chivukula:2014rka}. Finally, the experimental lower bound on the coloron mass is shown to be of order a few TeV by the Tevatron and LHC searches \cite{ColoronLim}.


\subsection{The Fermion Sector}\label{Ferm}

The extended strong interaction gauge sector allows the left- and the right-handed chiral eigenstates of the quarks (which are charged in the usual way under the electroweak interactions) to be charged under different $SU(3)_{i\,c}$ color gauge groups, giving rise to the interesting possibility of constructing a chiral theory of color \cite{Hill:1991at,Chivukula:1996yr,Frampton:1987dn,Bagger:1987fz,Martynov:2009en,Frampton:2009rk}.  Chiral charge assignments for the quarks under the extended color group will, subsequently, be reflected only in the quarks' chiral couplings to the colorons; their coupling to the gluons remains vector-like in nature, reproducing the ordinary QCD interactions regardless of the original charge assignments.

Such chiral color charge assignments for the quarks, however, can render the model anomalous in this sector \cite{Frampton:1987dn,Chivukula:2013xla,Frampton:1987ut,Cvetic:2012kv}, jeopardizing the overall consistency of the framework. The cancellation of such induced anomalies may be achieved by requiring the existence of new chiral spectator fermions \cite{Chivukula:2013xla,Cvetic:2012kv} whose chiral charges under the extended strong sector are the opposite of the chiral charges of the ordinary quarks. The number of spectator fermions required for anomaly cancellation depends on the details of the model under consideration. For instance, if all three generations of the ordinary quarks are chirally charged under the extended color gauge group, then three corresponding spectator fermion generations (carrying opposite chiral charges with respect to the quarks) are required to cancel the induced anomalies. On the other hand, if the chiral charge assignment of the third quark generation is opposite to those of the first two generations, only one additional spectator fermion generation (one up-like and one down-like spectator) is necessary. When all ordinary quarks are vectorially charged under the extended color interactions, no anomalies are induced and no spectator fermions are needed.  For further details, see Appendix~\ref{sec:appx-quark-charge}.

To achieve anomaly cancellation within the extended strong sector without introducing anomalies in other sectors, the introduced spectator fermions are conjectured to interact vectorially under the electroweak gauge group.\footnote{Spectators interacting chirally under the electroweak gauge group would require the addition of  ``lepton-like'' (color-neutral) spectator fermions to cancel anomalies introduced within the electroweak sector.}  Indeed, following \cite{Top-Coloron}, both the left- and the right-handed spectator generations are assumed to be doublets under the $SU(2)_{L}$ gauge group, while carrying a $U(1)_{Y}$ hypercharge $+1/6$. As a consequence, the spectator fermions carry the same electric charges as their corresponding quark counterparts; i.e., $+2/3$ for the up-like spectator and $-1/3$ for the down-like spectator.\footnote{The same assumption was made for the spectator electric charge assignments in the phenomenological studies of \cite{Chivukula:2013xka,Chivukula:2014rka}. As in \cite{Top-Coloron}, the hypercharge is normalized according to $Q=T_{3} + Y$.} A potential mixing between the spectators and the ordinary quarks was shown to be negligible, due to the presence of strong constraints on flavor-changing coloron couplings \cite{Top-Coloron}.

These anomaly-canceling spectator fermions obtain their masses via Yukawa interactions \cite{Chivukula:2013xka} with the bi-fundamental scalar $\Phi$ \eqref{Phi}:
\begin{equation}\label{Lferm}
- y_{Q} \tbrac{\bar{Q}^{k}_{R} \, \Phi\, Q^{k}_{L} + \bar{Q}^{k}_{L} \, \Phi^\dagger \, Q^{k}_{R}} \ ,
\end{equation}
with $Q^k_{L(R)}$ representing a left(right)-handed spectator doublet, and $k$ the generation index. Taking the Yukawa coupling $y_{Q}$ as a flavor-universal parameter for convenience,\footnote{In general, the Yukawa coupling can be a flavor matrix, giving rise to different masses for the individual spectator flavors.} one deduces the universal mass scale of the spectator flavors in terms of the singlet VEV:
\begin{equation}\label{MQ}
M_{Q} = \frac{y_{Q}}{\sqrt 6}\, v_{s} \ .
\end{equation}
While direct collider searches exclude spectator fermions with masses below only about 700 GeV, fits to precision electroweak observables raise the lower bound to a few TeV \cite{Spect}.

\vspace{1cm}

Summarizing our brief review, the renormalizable coloron model enlarges the SM color gauge group to $SU(3)_{1\, c} \times SU(3)_{2\, c}$ (c.f. \eqref{gaugestr}), spontaneously broken to the diagonal QCD subgroup. As a consequence, a set of massive color-octet vector bosons (colorons) are produced. The spontaneous symmetry breaking in the strong sector is facilitated by introducing additional (colored and color-neutral) scalar degrees of freedom.  Anomalies that would arise from the ordinary quarks' chiral charges under the extended color gauge group may be canceled by the addition of an appropriate number of new colored spectator fermions with opposite chiral charges.  As such, the model introduces eight free parameters, which may be expressed in terms of the physically relevant quantities \cite{Chivukula:2013xka,Chivukula:2014rka}:
\begin{equation}\label{freepar}
\cbrac{v_{s}, \sin \chi, m_{s}, m_{\cal A}, m_{G_{H}}, M_{C}, M_{Q}, N_Q} \ ,
\end{equation}
where $N_{Q}$ denotes the number of spectator fermion generations.

\section{Vacuum Stability and Triviality Analyses of the Scalar Potential}\label{Vacstab}

As described in Section~\ref{Prelim}, the renormalizable coloron model, with its extended strong sector, serves as an interesting candidate description of physics beyond the SM, introducing a variety of new vector, scalar and fermionic degrees of freedom into the theoretical framework. In this section, we examine the stability of the scalar potential \eqref{pot}, as well as the triviality of its couplings, up to an energy cutoff of 100~TeV. To this end, we construct the renormalization group (RG)-improved quartic interactions, by simultaneously solving the RG equations of the four scalar quartic couplings ($\lambda_{h}$, $\lambda_{m}$, $\lambda_{s}'$ and $\kappa_{s}$), and we determine their behavior as a function of the renormalization scale $\mu$. The stability of the scalar potential is, subsequently, guaranteed by requiring the conditions \eqref{stab} to be satisfied for the running couplings within the energy range of interest.\footnote{Note that the running of the $r_{\Delta}$ coupling does not interfere with the running of the quartic interactions (and, therefore, has no influence on the stability of the potential), since it is always accompanied by powers of the scalar VEV, $v_{s}$.} Furthermore, we demand that the running quartic couplings remain finite within the same energy range, by imposing an upper limit of $4\pi$ on their magnitudes; i.e., the triviality requirement should be fulfilled and no Landau poles should develop.

In this study, we distinguish four separate scenarios, according to the number of spectator fermion generations and the Yukawa or Dirac nature of their masses. If the spectator fermions obtain their masses via Yukawa interactions \eqref{Lferm} with the $\Phi$~scalar, then either 0, 1, or 3 generations of the spectator fermions may be present; we analyze each as a separate scenario.  However, if the theory is vectorial in nature---and thus anomaly free by construction--- another possibility still exists: one can include spectator fermions whose mass arises mainly from a Dirac mass term, with the Yukawa interaction \eqref{Lferm} negligible or entirely absent. Such a scenario with 1 generation of spectator fermions was introduced in \cite{Top-Coloron}, and we shall analyze the stability and triviality requirements for this scenario along with the three aforementioned cases.

In order to compute the running of the couplings as a function of the renormalization scale, it is necessary to obtain their $\beta$-functions within the context of the renormalizable coloron model. By solving the RG differential equation $\mu \, d \mathcal{C}/d \mu = \beta_{\mathcal C}$, the behavior of any of the running couplings~$\mathcal C$ as a function of the energy may, subsequently, be determined. The analytical expressions for the one-loop $\beta$-functions of the relevant couplings of the SM, the renormalizable coloron model containing spectators with Yukawa couplings, and the renormalizable coloron model containing spectators with a Dirac mass term are provided, respectively, in Appendices~\ref{SMBetaFun},~\ref{ReCoMBetaFun}, and \ref{DiracBetaFun}.

In the current treatment, we focus on capturing the essential physics of the renormalizable coloron model formalism, by considering one single relevant energy scale beyond the SM; namely, the singlet VEV, $v_{s}$. Taking $v_{s}$ as the characteristic energy scale where the spontaneous symmetry breaking of the extended color gauge group occurs, all the particle masses generated by the symmetry breaking may be approximated to reside in the vicinity of $v_{s}$. This is true for the new bosonic degrees of freedom, as well as the spectator fermions with the Yukawa interaction \eqref{Lferm}.

Assuming $v_s \gg m_{t}$, below the singlet VEV one may integrate out all of the non-SM degrees of freedom from the Lagrangian, recovering the ordinary SM and its $\beta$-functions as the low energy effective theory. Specifically, we start at $\mu = m_{t}$, with the $\overline{\text{MS}}$-scheme values of the gauge, the top Yukawa, and the Higgs\footnote{The normalization we use for $\lambda_{h}$ in \eqref{pot} differs by a factor of 6 from the conventional Higgs self coupling normalization in the literature. In addition, $g' = \sqrt{3/5} \, g_{1}$, where $g_{1}$ is the hypercharge coupling with the GUT normalization.} quartic couplings \cite{Zoller:2014cka}: 
\begin{equation}
g(m_{t}) = 0.6483\,, \quad g'(m_{t}) = \sqrt{3/5}\times 0.3587\, , \quad g_{s}(m_{t}) = 1.1671\, , \quad y_{t}(m_{t}) = 0.9369\, , \quad\lambda_{h}(m_{t}) = 6\times0.1272\ .
\end{equation}
As the renormalization scale is increased, the new degrees of freedom associated with the renormalizable coloron model become kinematically active above $v_s$, and these affect the $\beta$-functions and the running of the couplings. Therefore, the appropriate matching between the low and high energy running of the different couplings must occur at the threshold $v_s$.\footnote{By integrating out all states simultaneously, we neglect one-loop corrections proportional to logarithms of ratios of masses relative to $v_s$, e.g. proportional to $\log(M^2/v^2_s)/16\pi^2$.} In the following paragraphs, we discuss the separate behaviors of the gauge, fermion, and scalar sectors both below and above this threshold.

At low energies, $\mu < v_s$, the strong gauge sector resides in the broken phase and acts like ordinary QCD, interacting via the massless gluons with the running strong coupling $g_{s}$. However, once the threshold is reached, the colorons become kinematically active and fully participate in the interactions; the structure of the strong gauge sector is restored to the unbroken phase $SU(3)_{1\, c} \times SU(3)_{2\, c}$. The strong coupling $g_{s}$ is, subsequently, replaced by the two running couplings of the extended color gauge group, $g_{s_{1}}$ and $g_{s_{2}}$, with the matching conditions at the threshold $\mu = v_s$ given by \eqref{gs12Match}. Note that the matching conditions \eqref{gs12Match} result in different shifts for the staring points of the two $g_{s_{i}}$ couplings, with respect to the running value of the QCD coupling at that scale, $g_{s}(v_s)$.

Spectator fermions with Yukawa interactions described by \eqref{Lferm} do not contribute below $\mu < v_s$, and become kinematically accessible only above the threshold. At $\mu \geq v_s$, their Yukawa coupling and its $\beta$-function is activated, with the starting point given by \eqref{MQ}. For spectator fermions with a Dirac mass term, the kinematic threshold above which these fermions start to participate is, instead, set by their Dirac mass $M_{Q}$, which is independent of the symmetry breaking scale $v_{s}$. In this scenario, the fermionic threshold, $M_{Q}$, should be treated as a free parameter separate from the bosonic threshold $v_s$. Note also that, as mentioned in Section~\ref{Ferm}, mixing between the ordinary quarks and the spectator fermions is negligible \cite{Top-Coloron}, and their Yukawa couplings remain independent of one another. 

The scalar sector also has distinctive behavior in different energy regimes. Below $\mu < v_s$, one may integrate out the $(3,\bar 3)$ $\Phi$~scalar altogether from the potential \eqref{pot} by completing the square. This reproduces the SM scalar potential with the effective Higgs coupling as the sole scalar interaction:
\begin{equation}\label{lheff}
\lambda_{h}^{\text{eff}}(\mu < v_s) \equiv \lambda_{h} - 9 \frac{\lambda_{m}^{2}}{\lambda_{s}'} \ ,
\end{equation}
which arises due to the mixing between the SM Higgs doublet $\phi$ and the complex $\Phi$~scalar. Once the threshold is reached, $\mu \geq v_s$, the $\Phi$~scalar and its additional spin-0 degrees of freedom become kinematically accessible and the remaining (quartic) couplings participate within the scalar interactions. The $\beta$-functions of the four scalar quartic couplings are, at this point, coupled to one another and must be solved simultaneously. The matching conditions for these couplings are, at the scale $\mu = v_s$, retrieved from the expressions in \eqref{scalarMatch} that represent the starting point for the quartic couplings $\lambda_{m}$, $\lambda_{s}'$, and $\kappa_{s}$, which are absent at low energies.  The transition of the low energy $\lambda_{h}^{\text{eff}}$ into the high energy $\lambda_{h}$ is accompanied by a shift at the scale $v_s$, due to the mixing quartic coupling (c.f. \eqref{lheff}).

In the following section, we identify the viable region of the free parameter space, in which the stability and triviality conditions are satisfied for the four scenarios containing different configurations of spectator fermions. We illustrate our results using exclusion plots.

\section{Results and Discussion}\label{Resl}

Having discussed the formal structure of the model and set up the vacuum stability and triviality requirements in the previous sections, we now proceed to determine the viable regions of the parameter space where these conditions can be fulfilled. Specifically, in this study we are interested in imposing the vacuum stability of the potential and triviality conditions up to an energy of 100~TeV; in other words, we demand that the conditions \eqref{stab} should be satisfied for all of the four running scalar quartic couplings within this energy range, and that none of the couplings should develop a Landau pole. We will investigate separately the three scenarios of 0, 1, or 3 generations of the spectator fermions having Yukawa interactions \eqref{Lferm} with the $\Phi$~scalar, as well as the scenario of 1 spectator fermion generation with Dirac masses.

The obtained viable regions of the parameter space are exhibited in $\sin\chi - m_s$ exclusion plots, taking into account both the positive and negative values of the mixing angle---corresponding to the sign of $\lambda_{m}$, signifying an attractive or repulsive interaction between the scalars $\phi$ and $\Phi$. Furthermore, the plots incorporate the theoretical and the 95\%~C.L. experimental constraints from demanding the global minimum of the potential to coincide with the scalar VEVs (c.f. \eqref{rdel} and \eqref{rdelthresh}), unitarity, electroweak precision tests, and the LHC direct measurements of the Higgs boson couplings, which were previously analyzed and derived in \cite{Chivukula:2013xka}.

\subsection{Scenario with No Spectator Fermions ($N_{Q}=0$)}\label{NQ0}

In this scenario, the ordinary quarks are vectorially charged under the extended color group $SU(3)_{1\, c} \times SU(3)_{2\, c}$; i.e., both the left- and right-handed eigenstates of each chiral quark flavor are charged under the same gauge group $SU(3)_{i\, c}$. The scenario is, hence, anomaly-free by construction and no additional spectator fermions with Yukawa couplings to the $\Phi$~scalar are required for anomaly cancellation purposes. For concreteness, this scenario is represented by vectorially charging all three quark generations under $SU(3)_{1\, c}$ by convention.\footnote{This choice is, however, immaterial for the overall phenomenological results.} The $\beta$-functions of the running couplings, above the singlet VEV scale $v_s$, are given in Appendix~\ref{ReCoMBetaFun} by setting $N_{Q}=0$ and $y_{Q}=0$. None of the couplings, therefore, receives any contribution from the spectator fermions in their running.

\begin{figure}
\includegraphics[width=.49\textwidth]{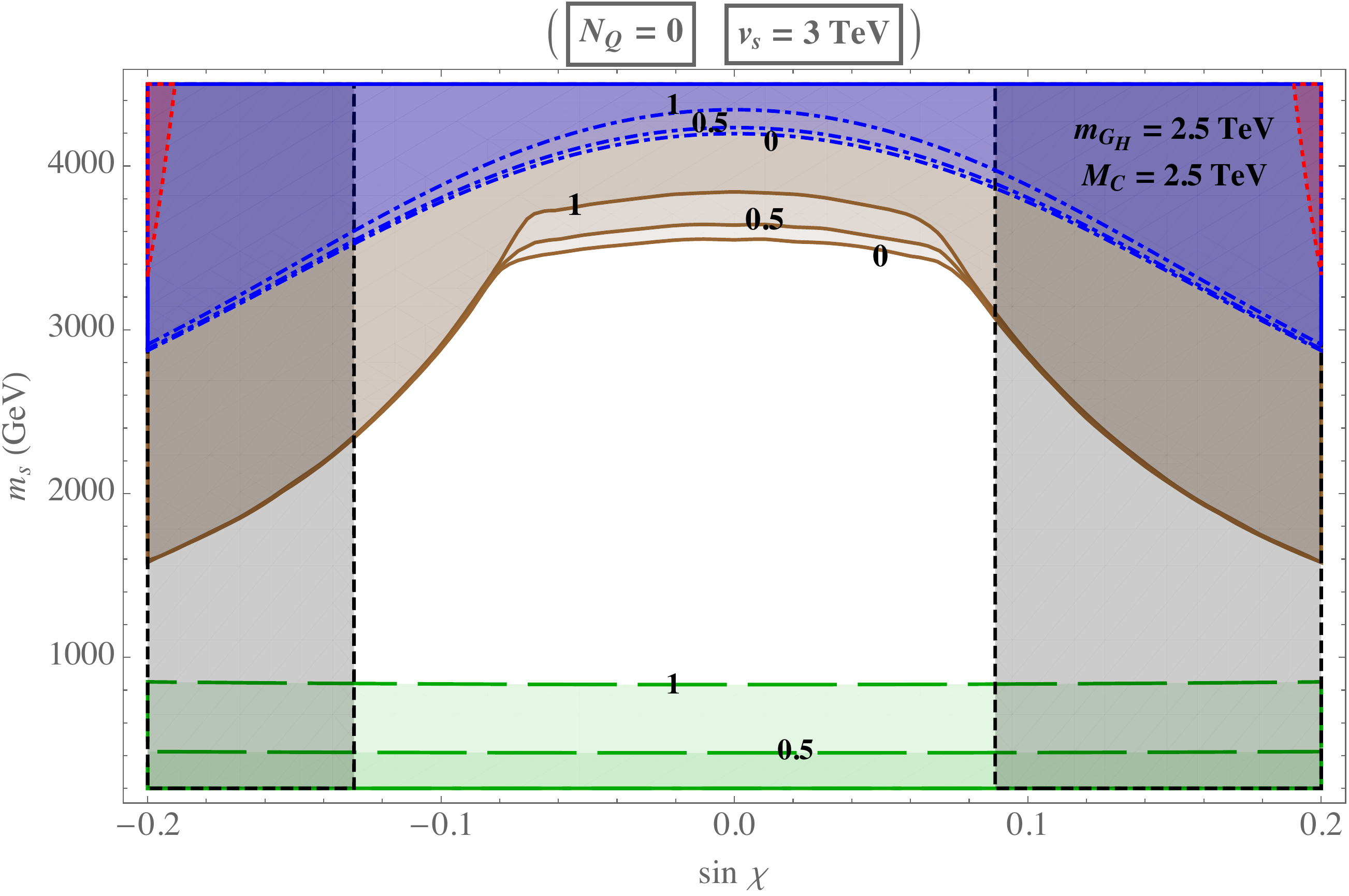}
\includegraphics[width=.49\textwidth]{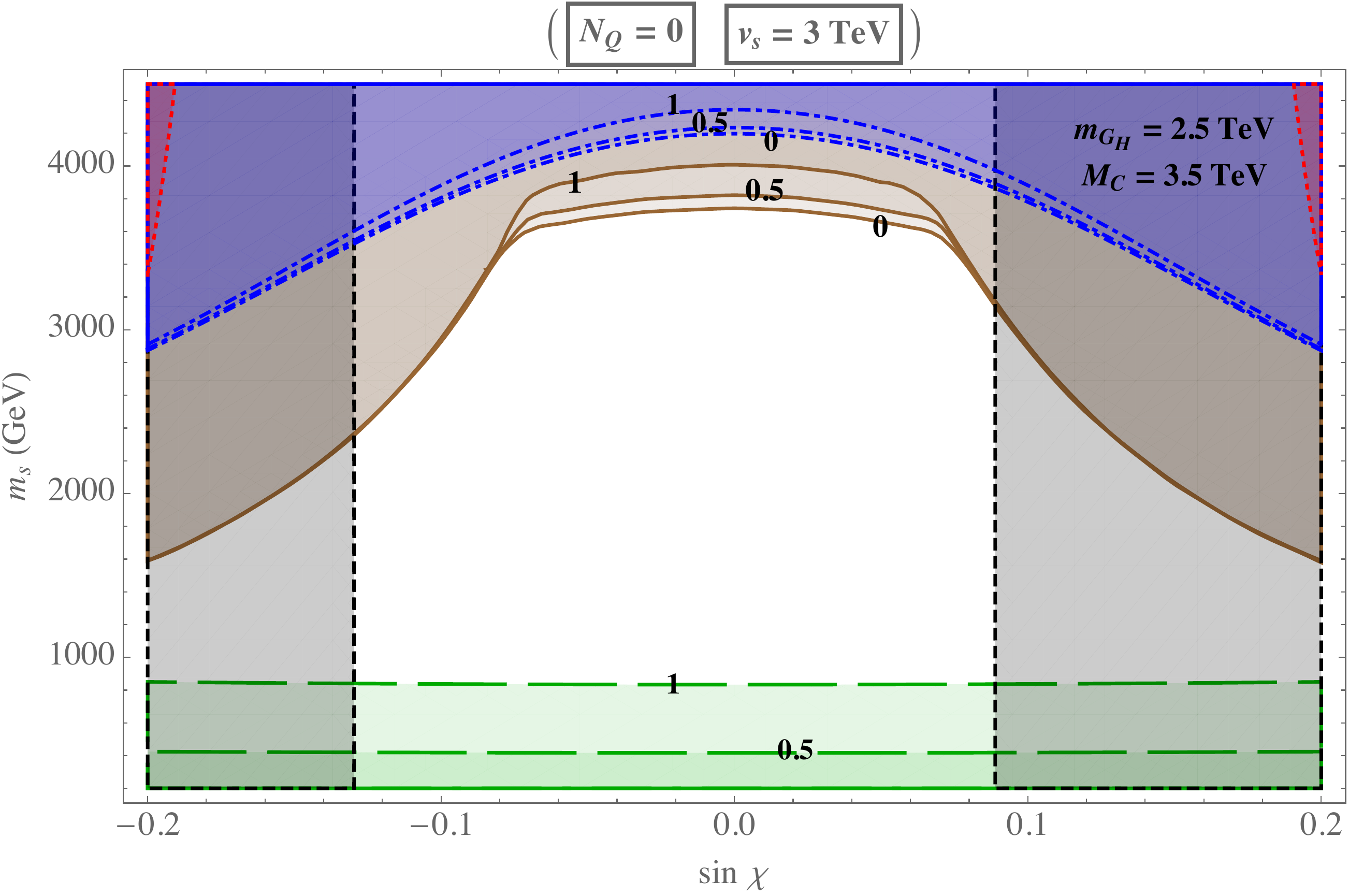}
\includegraphics[width=.49\textwidth]{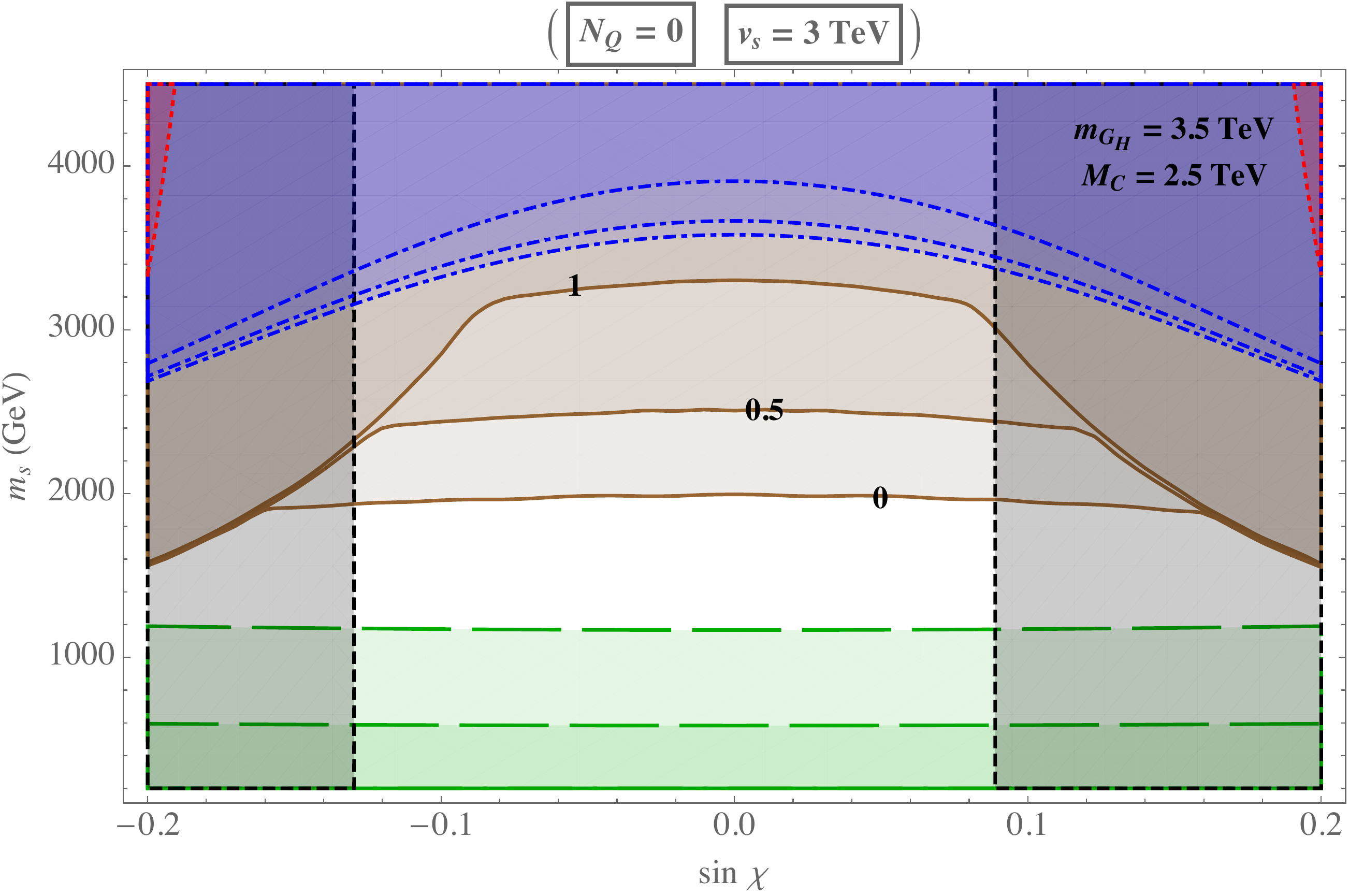}
\includegraphics[width=.49\textwidth]{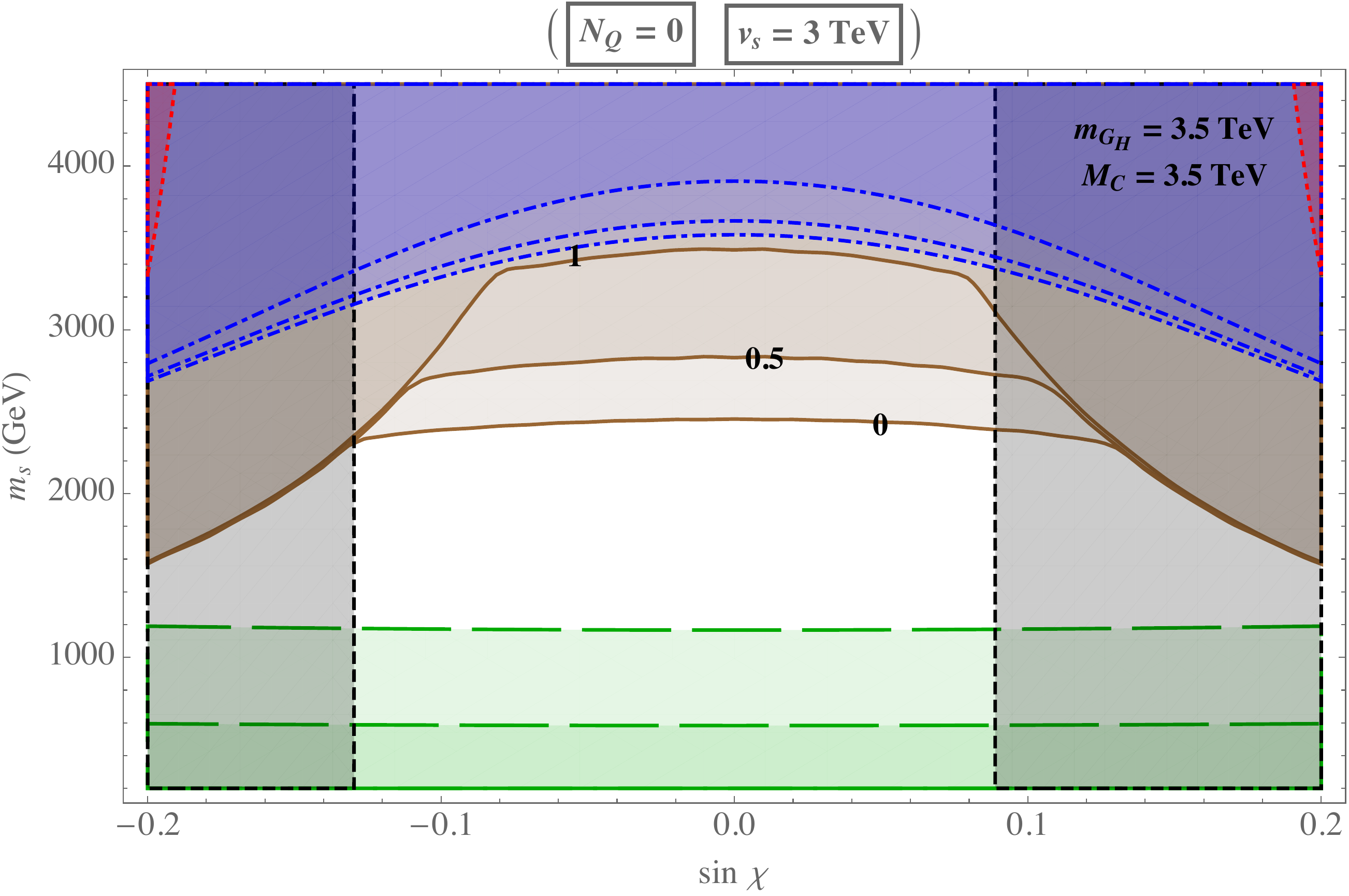}
\caption{The $\sin\chi - m_s$ exclusion plots with no spectator fermions, $N_{Q}=0$, for a singlet VEV $v_{s}= 3$~TeV. The panels display benchmark values of $m_{G_H}, M_C \in \cbrac{2.5, 3.5}$~TeV, so that one can observe the impact of having each mass lying below or above $v_s$.  The constraints from considering the global minimum of the potential coinciding with the VEVs (long-dashed green line), unitarity (dot-dashed blue line), electroweak precision tests (dotted red line), direct Higgs couplings' measurements by the LHC (short-dashed black line), and vacuum stability and triviality up to 100~TeV (solid brown line) are exhibited.  The enumerated curves correspond to several values of $n \equiv m_{\mathcal A}/m_{G_H}$ between 0 and 1. All colored regions are excluded.}
\label{resultNQ0vs3}
\end{figure}

The scenario contains six free parameters: $v_{s}$ and $\sin\chi$, as well as the bosonic masses $m_s$, $m_{\mathcal A}$, $m_{G_H}$ and $M_C$. The panels in Fig.~\ref{resultNQ0vs3} display the $\sin\chi - m_s$ exclusion plots for a singlet VEV, $v_s=3$~TeV. The $M_C$ and $m_{G_H}$ values are varied from below $v_s$ to above it in each row and column, respectively. Within each panel, the exclusion curves corresponding to three representative values of $n \equiv m_{\mathcal A} / m_{G_H}$ are superimposed, $n= 0, 0.5$, and 1.\footnote{Note that the upper value of $m_{\mathcal A}$ is given by \eqref{mAcond}, resulting in $0 \leq n \lesssim 1.2 $.} The bounds are obtained by imposing the vacuum stability and triviality conditions up to 100~TeV, as well as the previously analyzed \cite{Chivukula:2013xka} theoretical and experimental constraints. 

\begin{figure}
\includegraphics[width=.49\textwidth]{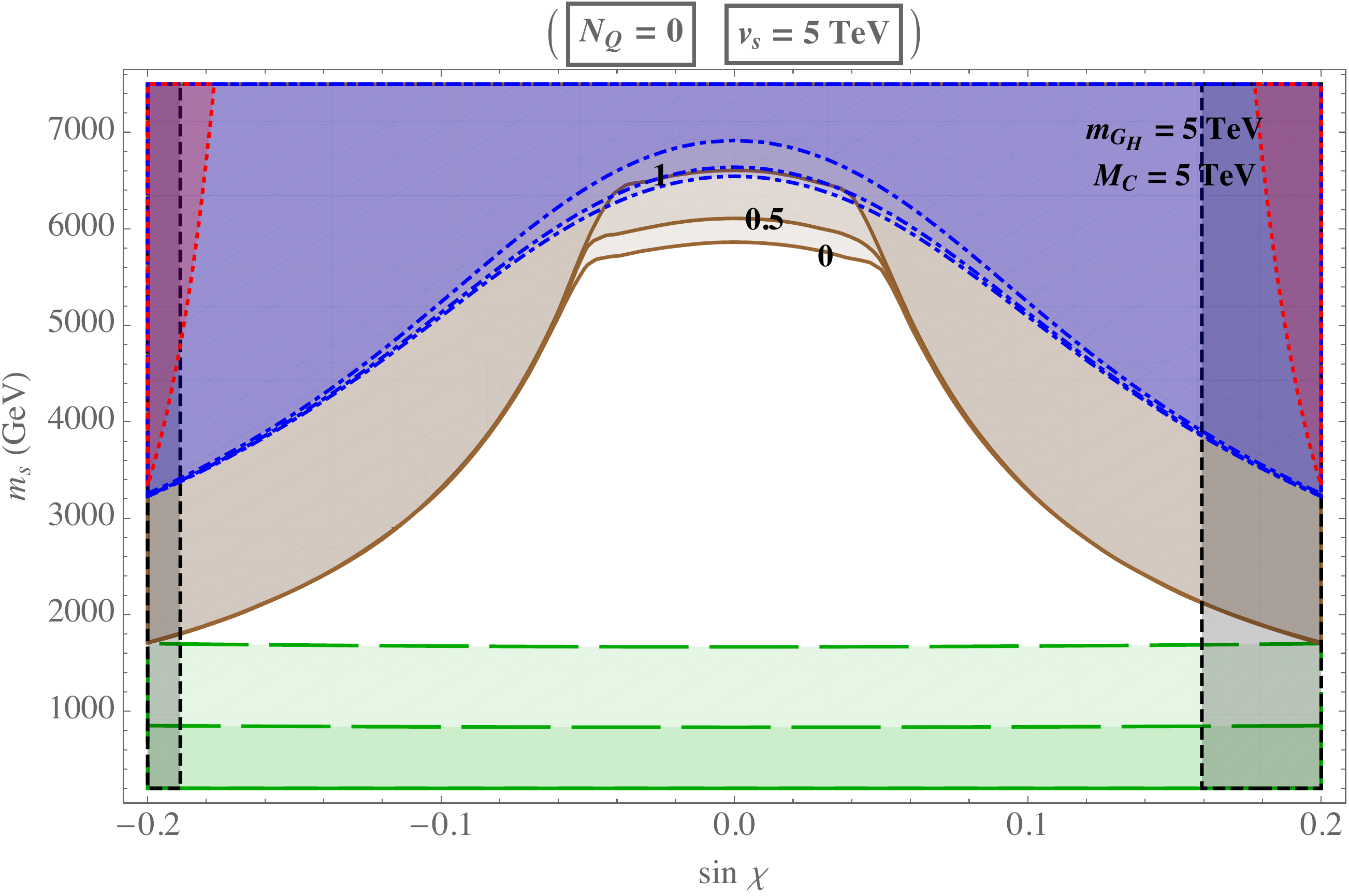}
\includegraphics[width=.49\textwidth]{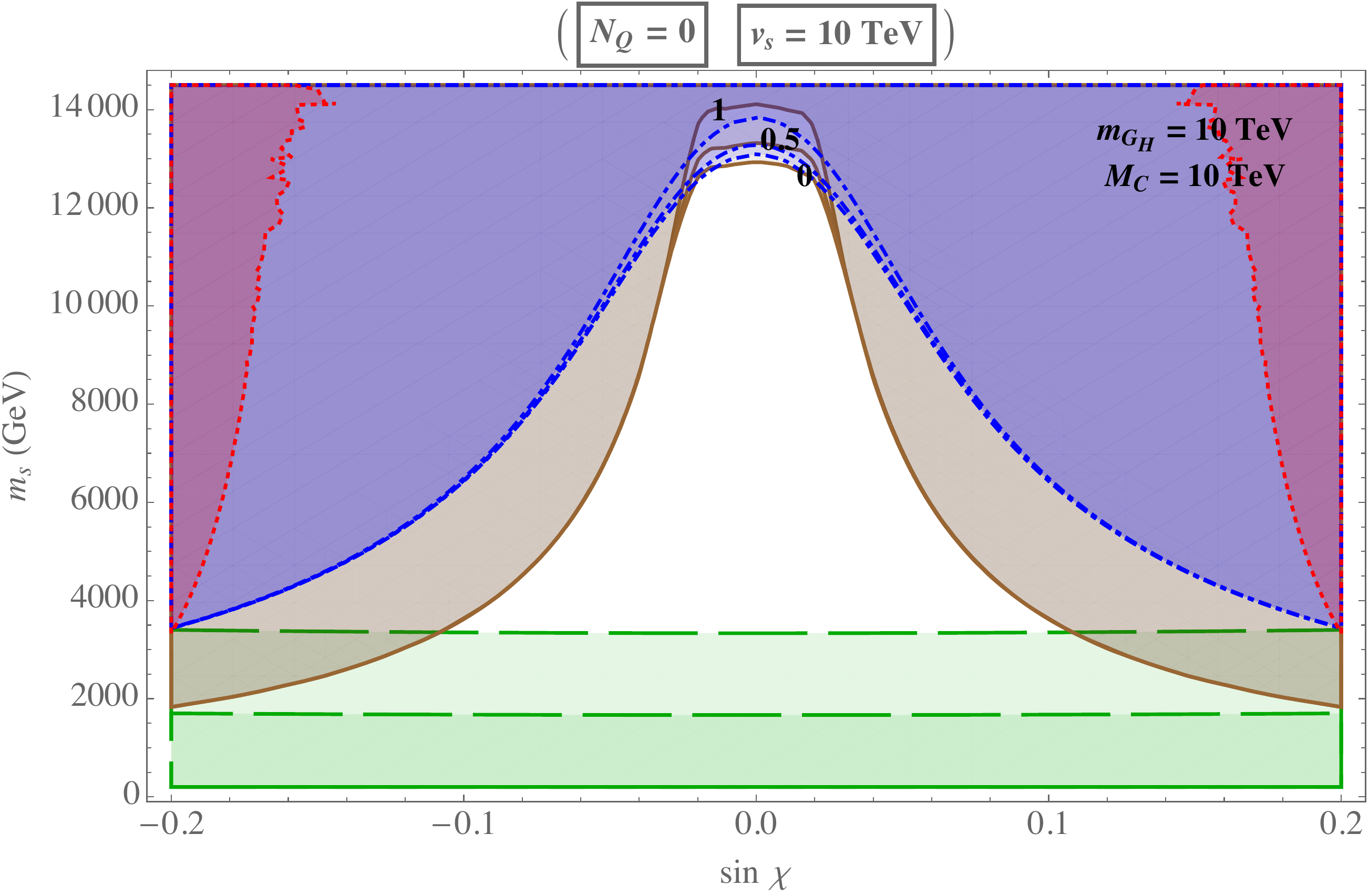}
\caption{The $\sin\chi - m_s$ exclusion plots with no spectator fermions, $N_{Q}=0$, for singlet VEVs $v_{s}= 5,10$~TeV, and $m_{G_H},\, M_C \, \sim\, v_s$. Once more, several $n \equiv m_{\mathcal A}/m_{G_H}$ curves between 0 and 1 are displayed. All colored regions are excluded. Note that as $v_s$ increases, the unitarity bounds (corresponding, from bottom to top, to $n=0,0.5,1$) restrict $\sin\chi$ more strongly, the direct Higgs constraints restrict $\sin\chi$ less strongly, and the regions allowed by electroweak precision tests, vacuum stability and triviality become both more constrained in $\sin\chi$ and less constrained in $m_s$. Once more, the lower bounds (corresponding, from bottom to top, to $n=0.5,1$) are due to demanding the global minimum of the potential to coincided with the VEVs. (see the caption of Fig.~\ref{resultNQ0vs3} for further details)}
\label{resultNQ0vs510}
\end{figure}

In these exclusion plots, the upper bound on $m_s$ in different regions, due to the vacuum stability and triviality constraints, is determined by the following two competing effects:
\begin{itemize}
\item At relatively small values of $|\sin \chi|$, the quartic coupling $\lambda_{s}'$ develops a Landau pole for large values of $m_s$ before the renormalization scale has reached the 100~TeV cutoff. This process, however, depends only weakly on the precise value of the mixing angle.
\item As the absolute value of the mixing angle increases, the SM Higgs coupling $\lambda_{h}$ receives larger contributions at the threshold $v_s$ (c.f. \eqref{scalarMatch}) as well as to its running from the mixing coupling $|\lambda_{m}|$, and develops a Landau pole before the cutoff is reached. This process is much more strongly dependent on the mixing angle. 
\end{itemize}
The upper bound on $m_s$ is, thus, determined by the triviality condition in this scenario. These two competing effects explain the origin of the sudden drop in the upper value of $m_s$ as a function of $|\sin \chi|$: the observed ``kinks'' in Fig.~\ref{resultNQ0vs3}. The much stronger dependence of $\lambda_{h}$ on the mixing angle, as compared with $\lambda_{s}'$, is encoded in their $\beta$-functions \eqref{bquartic}, with the former receiving a much larger contribution in its running from $\lambda_{m}$ (due to the larger coefficient) than the latter. Furthermore, the $\beta$-function of $\lambda_{m}$ reveals different running behaviors for the positive and negative values of this coupling, although with very moderate effects on the viable parameter space (the negative values of $\lambda_{m}$ slightly mitigate the constraints on the viable parameter space).

\begin{figure}
\includegraphics[width=.49\textwidth]{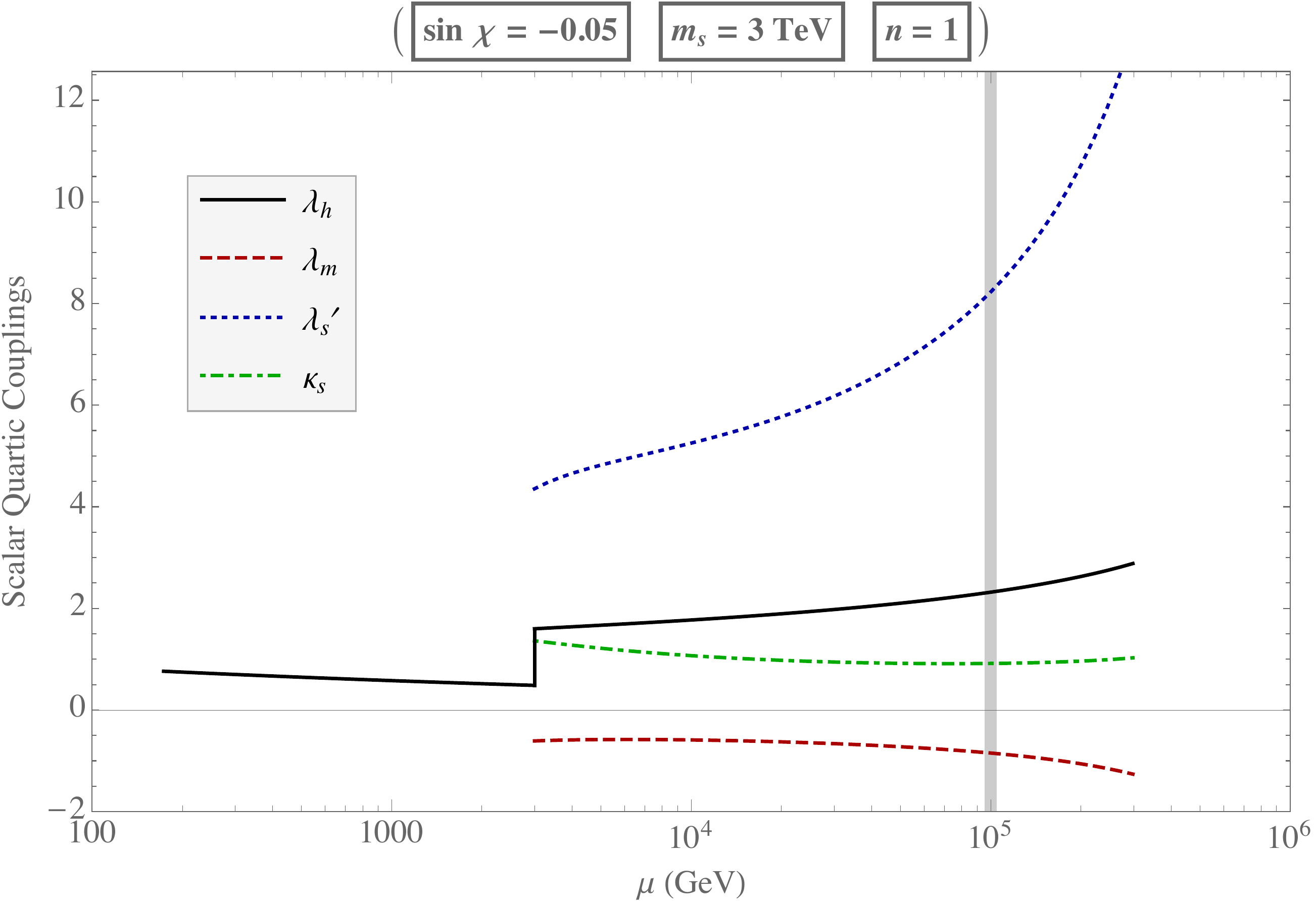}
\includegraphics[width=.49\textwidth]{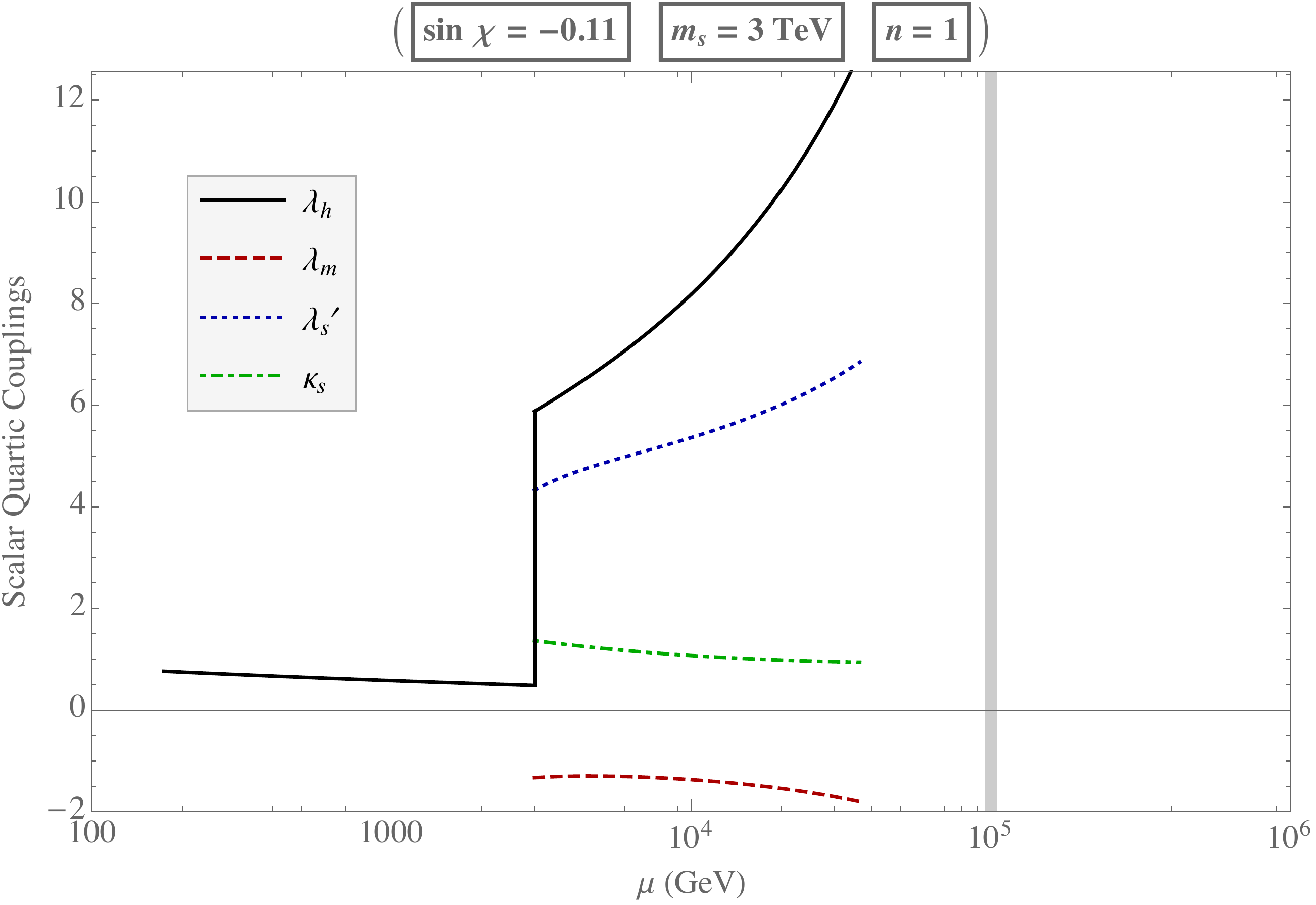}
\includegraphics[width=.49\textwidth]{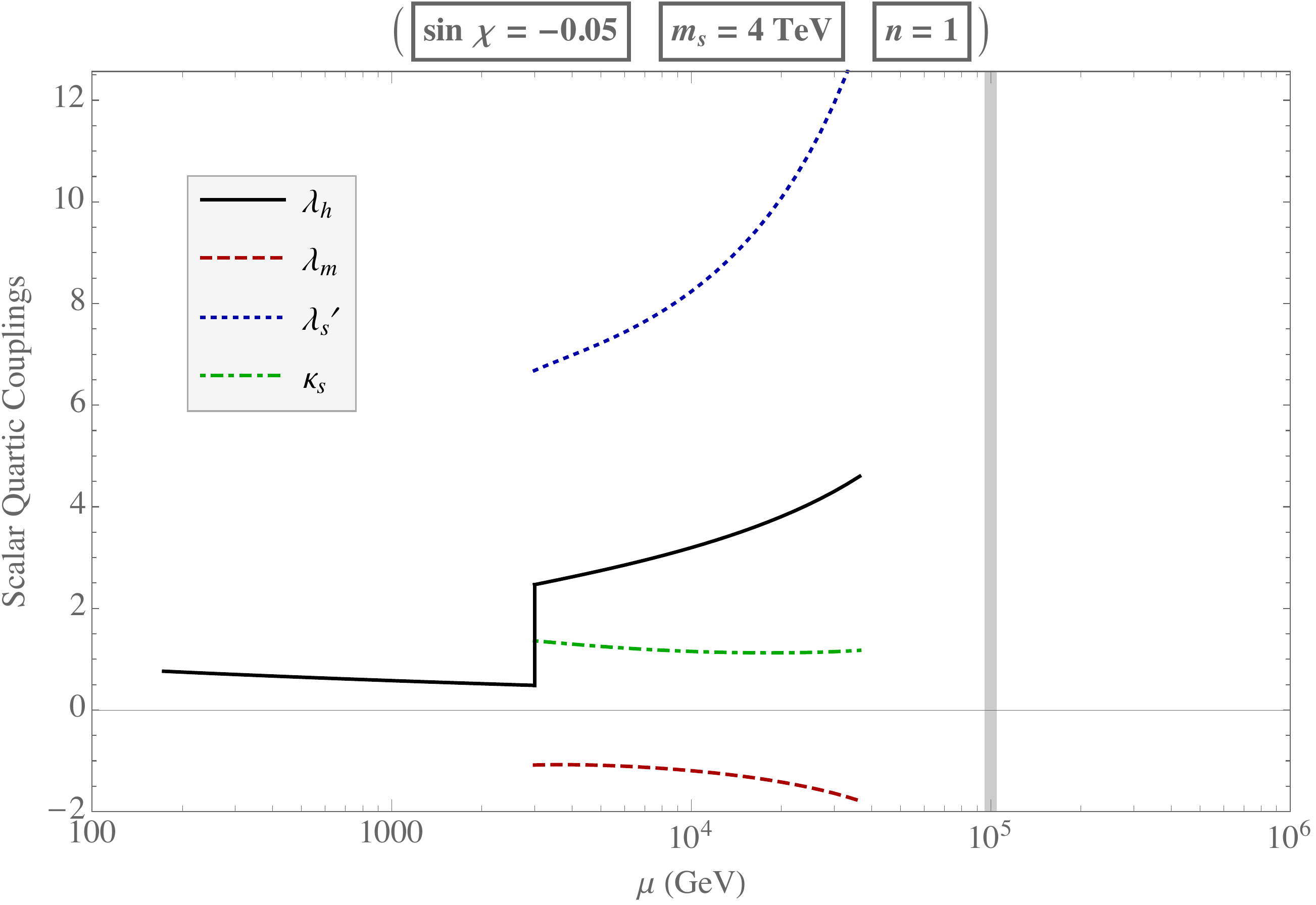}
\includegraphics[width=.49\textwidth]{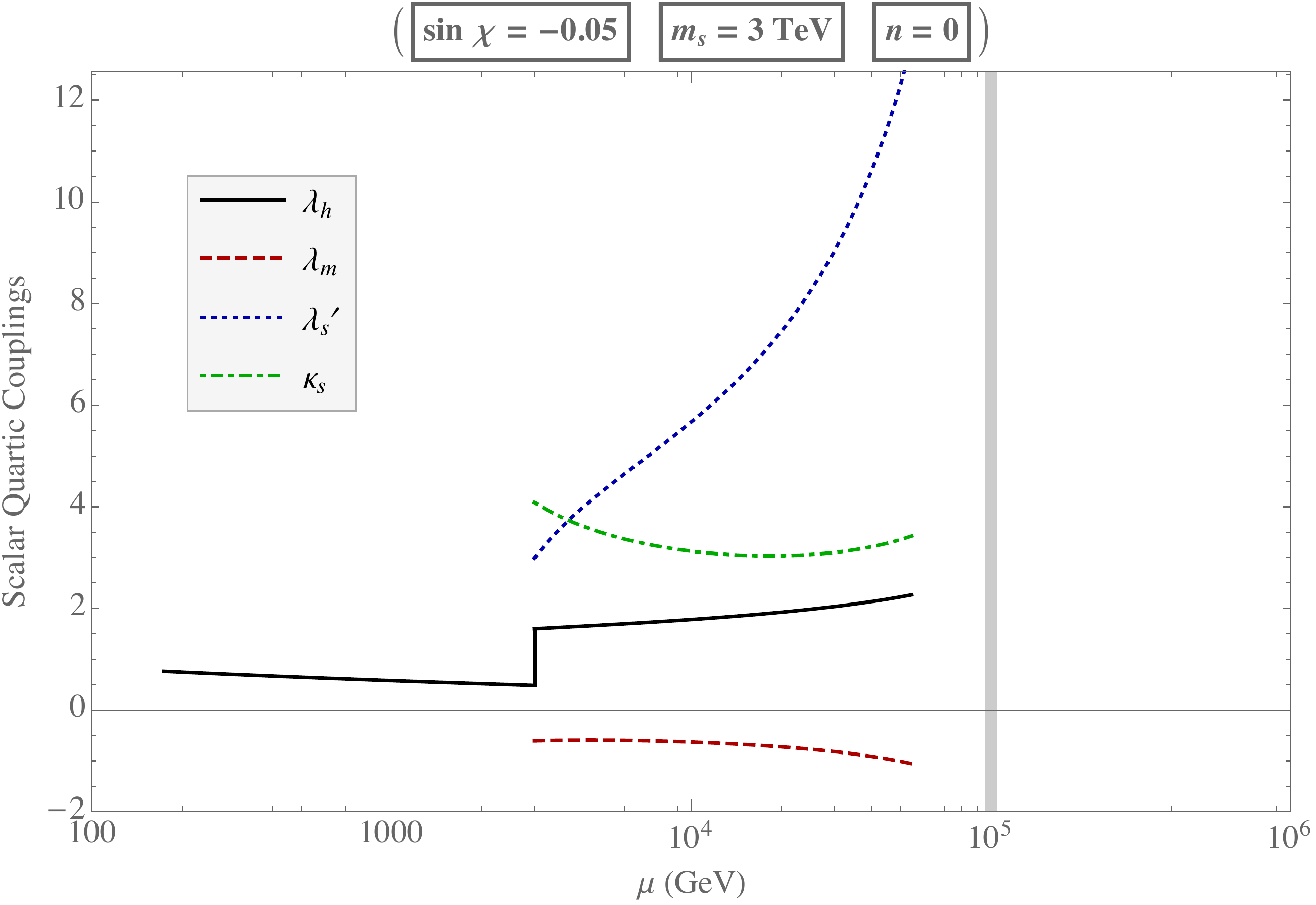}
\includegraphics[width=.5\textwidth]{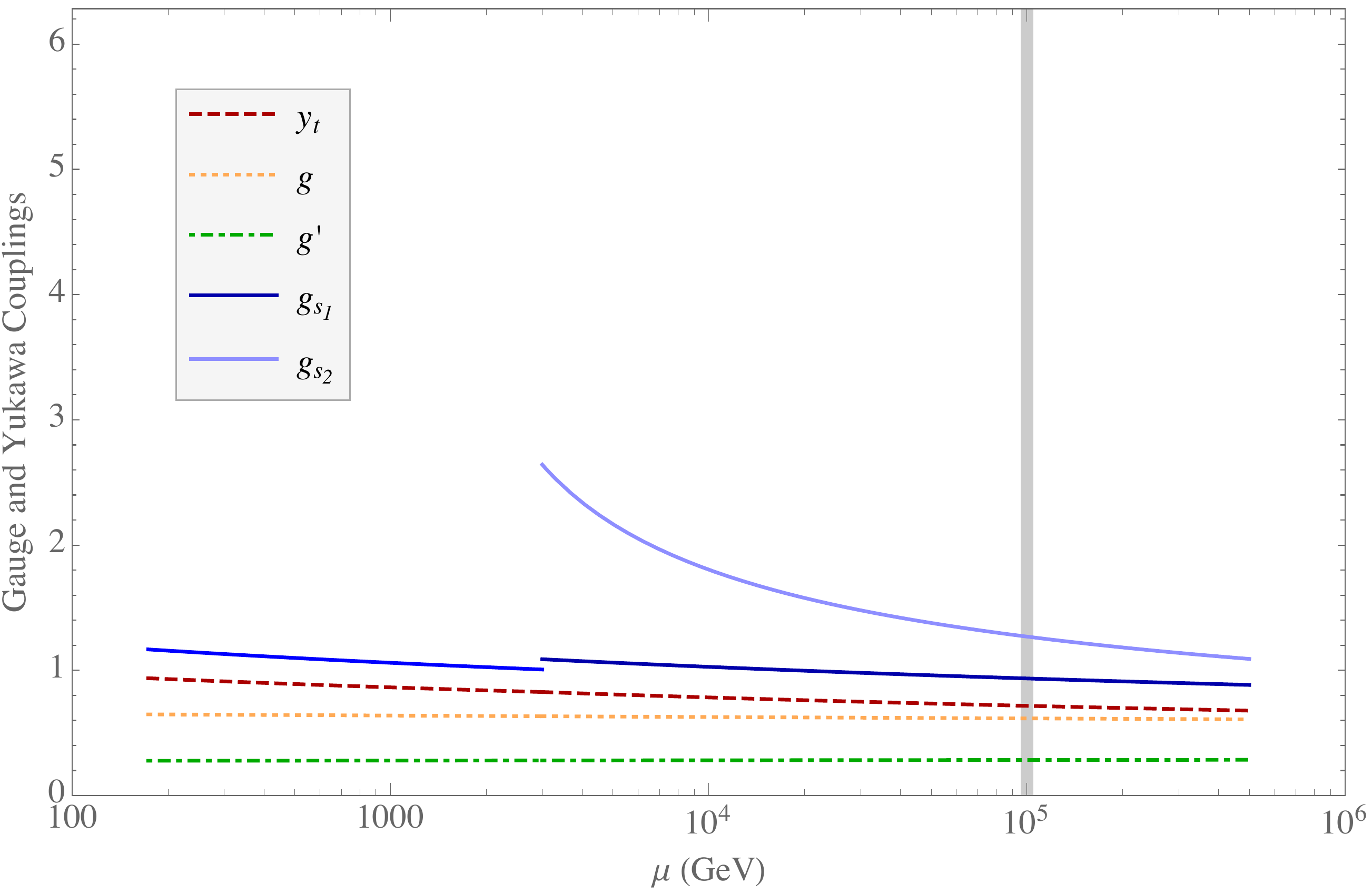}
\caption{The running of the scalar quartic couplings (top and middle rows) and the gauge and Yukawa couplings (bottom panel) as a function of the renormalization scale $\mu$, for $N_{Q}=0$, $v_{s}=3$~TeV, $m_{G_H} = M_C = 3.5$~TeV (bottom-right panel of Fig.~\ref{resultNQ0vs3}). The panels in the top and middle rows correspond to different regions of the $\sin\chi - m_s$ parameter space: the top-left panel illustrates a point in the allowed region for a heavy pseudoscalar, the top-right panel corresponds to a point excluded because $\sin\chi$ is too large, and the middle-left panel corresponds to a point excluded because $m_s$ is too large. The middle-right panel shows a point that is allowed for a heavy pseudoscalar (as in the top-left panel) but becomes excluded when the pseudoscalar is light. The vertical solid line indicates the imposed 100~TeV cutoff, and a Landau pole is indicated by a coupling that exceeds $4\pi$; in the top-right panel, $\lambda_h$ suffers a Landau pole below 100 TeV, while in the middle row it is $\lambda_s'$ that does so. The flow of the gauge and Yukawa couplings (bottom panel) is the same in all panels, since they do not depend on the mixing angle or the (pseudo)scalar masses. Note that the Higgs quartic coupling, $\lambda_{h}$, receives a shift at the threshold $v_s$, whereas the strong coupling $g_{s}$ is replaced by the two extended gauge group couplings, $g_{s_{1}}$ and $g_{s_{2}}$, at the same threshold.}
\label{couplingsNQ0}
\end{figure}

As mentioned, within each panel, the mass of the pseudoscalar, $m_{\mathcal A}$, is varied from 0 to near its formal upper bound \eqref{mAcond}. According to \eqref{scalarMatch}, a larger $m_{\mathcal A}$ introduces, at the threshold, a \textit{smaller} starting point for the $\kappa_s$ quartic coupling, to which the $\beta$-function of $\lambda_{s}'$ is quadratically sensitive with a sizable coefficient. Hence, a larger $m_{\mathcal A}$ mitigates the positive $\kappa_s$ contribution to the running of the $\lambda_{s}'$ coupling, allowing for an increased $m_s$~upper bound at small mixing angles before the Landau pole is reached.\footnote{At the same time, a larger $m_{\mathcal A}$ directly induces a \textit{larger} threshold starting point for $\lambda_{s}'$ by \eqref{scalarMatch}; however, this enhancing effect is more than compensated by the corresponding decrease of the $\kappa_s$ contribution to the $\lambda_{s}'$ running, and the resulting viable parameter space is enlarged (see Fig.~\ref{couplingsNQ0}).} In contrast, the $m_s$~upper bound at larger mixing angles, caused when $\lambda_{h}$ developes a Landau pole, is independent of the value of the pseudoscalar mass $m_{\mathcal A}$, as expected. A larger scalar color-octet mass $m_{G_H}$ (lower plots in Fig.~\ref{resultNQ0vs3}), on the other hand, forces a greater $\kappa_s$ initial value at the threshold and an enhanced positive contribution to the running of $\lambda_{s}'$, which results in the latter's developing a Landau pole at smaller $m_s$. In this case, an additional change in $m_{\mathcal A}$ has a much more pronounced effect on the $m_s$~upper bound and the resulting viable parameter space.

Finally, one observes that a (slightly) larger coloron mass (right-hand panels in Fig.~\ref{resultNQ0vs3}) slightly alleviates the Landau pole constraints. This is attributed to the fact that, within the scalar $\beta$-functions (c.f. \eqref{bquartic}), the two $g_{s_i}$ extended gauge couplings (whose threshold values depend on the coloron mass by \eqref{gs12Match}) contribute according to two competing terms: a positive quartic term $\propto  +\pbrac{g^{2}_{s_{1}} + g^{2}_{s_{2}}}^{2}$ and a negative quadratic term $\propto - \pbrac{g^{2}_{s_{1}} + g^{2}_{s_{2}}}\lambda_{\text{quartic}}$. For coloron masses not far away from the singlet VEV, $M_C \sim v_s$, the values of the two $g_{s_i}$ couplings start relatively small, and the negative quadratic term (multiplied by the potentially sizable scalar quartic coupling) takes the upper hand. This in turn, causes a taming of the positive running of the scalar quartic coupling, postponing its Landau pole. Such behavior, nevertheless, reverses for larger coloron masses lying farther away from the singlet VEV, as the positive quartic gauge term is dominant for large $g_{s_i}$ threshold values.

The dependence of the viable region of the parameter space on the singlet VEV is further illustrated in Fig.~\ref{resultNQ0vs510} for two larger values, $v_s = 5$~and~10~TeV. We illustrate the benchmark case where $M_C,\, m_{G_H}\, \sim \, v_s$, since the impact of varying the color octet states' masses away from $v_s$ remains similar to that analyzed for $v_s = 3$~TeV in Fig.~\ref{resultNQ0vs3}. One notes that a larger $v_{s}$ necessitates larger $m_s$, which in turn induces a larger contribution to $\lambda_{h}$ at the threshold, for a given value of the mixing angle, and leads to the development of a Landau pole more quickly. Therefore, the viable region is dominated by smaller values of the mixing angle for larger singlet VEVs.

In order to explicitly demonstrate these findings, we plot the running of the couplings as a function of the renormalization scale in Fig.~\ref{couplingsNQ0}, taking as an example different locations within the parameter space at $v_{s}=3$~TeV, $m_{G_H} = M_C = 3.5$~TeV (bottom-right panel of Fig.~\ref{resultNQ0vs3}). A representative value within the allowed region of the parameter space for a heavy pseudoscalar is selected for the top-left panel of Fig.~\ref{couplingsNQ0}. One observes that, for this choice of the free parameters, all couplings behave as expected below the 100~TeV cutoff; i.e., the stability of the potential \eqref{stab} is guaranteed and no Landau poles are encountered. The top-right panel exhibits the situation where the mixing angle has been increased until it falls outside the allowed region. In this case, the Higgs quartic coupling, $\lambda_{h}$, receives unacceptably large contributions from the mixing coupling, $\lambda_{m}$, at the threshold as well as in its running, and develops a Landau pole before reaching the cutoff, excluding this region. In contrast, the situation where $m_s$ has been increased to fall outside the allowed region is displayed in the middle-left panel of the same figure. Here, the larger value of $m_s$ induces larger contributions to the running of $\lambda_{s}'$ via positive bosonic coefficients in its $\beta$-function, as well as large contributions to its starting point at the threshold (c.f. \eqref{scalarMatch}). As a consequence, large $m_s$~values lead to a Landau pole for this quartic scalar coupling, excluding the corresponding region of the parameter space. The middle-right panel illustrates the situation where a region that is allowed when the pseudoscalar is heavy ($n \sim 1$, top-left panel) becomes excluded if the pseudoscalar is light ($n \sim 0$). As previously explained, this happens since a light pseudoscalar implies a larger starting point for the $\kappa_s$~coupling, which then unacceptably enhances the running of $\lambda_{s}'$, causing the latter to develop a Landau pole despite its lower threshold value.

\subsection{Scenario with 1 Spectator Fermion Generation with Yukawa Interactions ($N_{Q}=1$)}\label{NQ1}

When the chiral eigenstates of the third generation of ordinary quarks are charged under the extended color gauge group in the opposite way to those of the first two generations, then 1 generation of the spectator fermions, with a Yukawa coupling to the $\Phi$~scalar, is necessary to cancel the induced anomalies. The $\beta$-functions of the couplings, above the threshold $v_s$, are given in  Appendix~\ref{ReCoMBetaFun} by setting $N_{Q}=1$. The spectator Yukawa coupling, $y_{Q}$, is fully active, and contributes to the relevant scalar coupling $\beta$-functions.

\begin{figure}
\includegraphics[width=.49\textwidth]{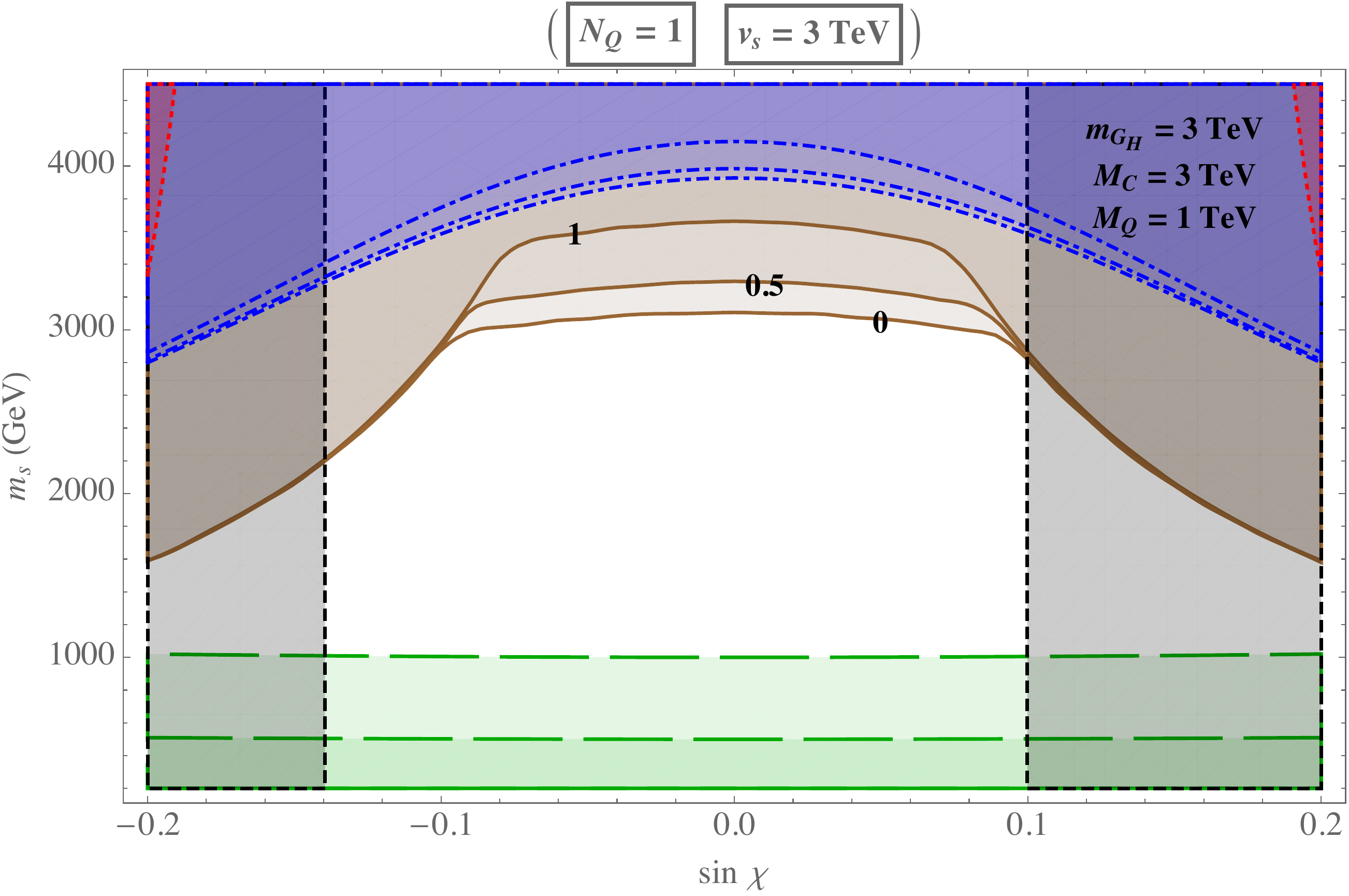}
\includegraphics[width=.49\textwidth]{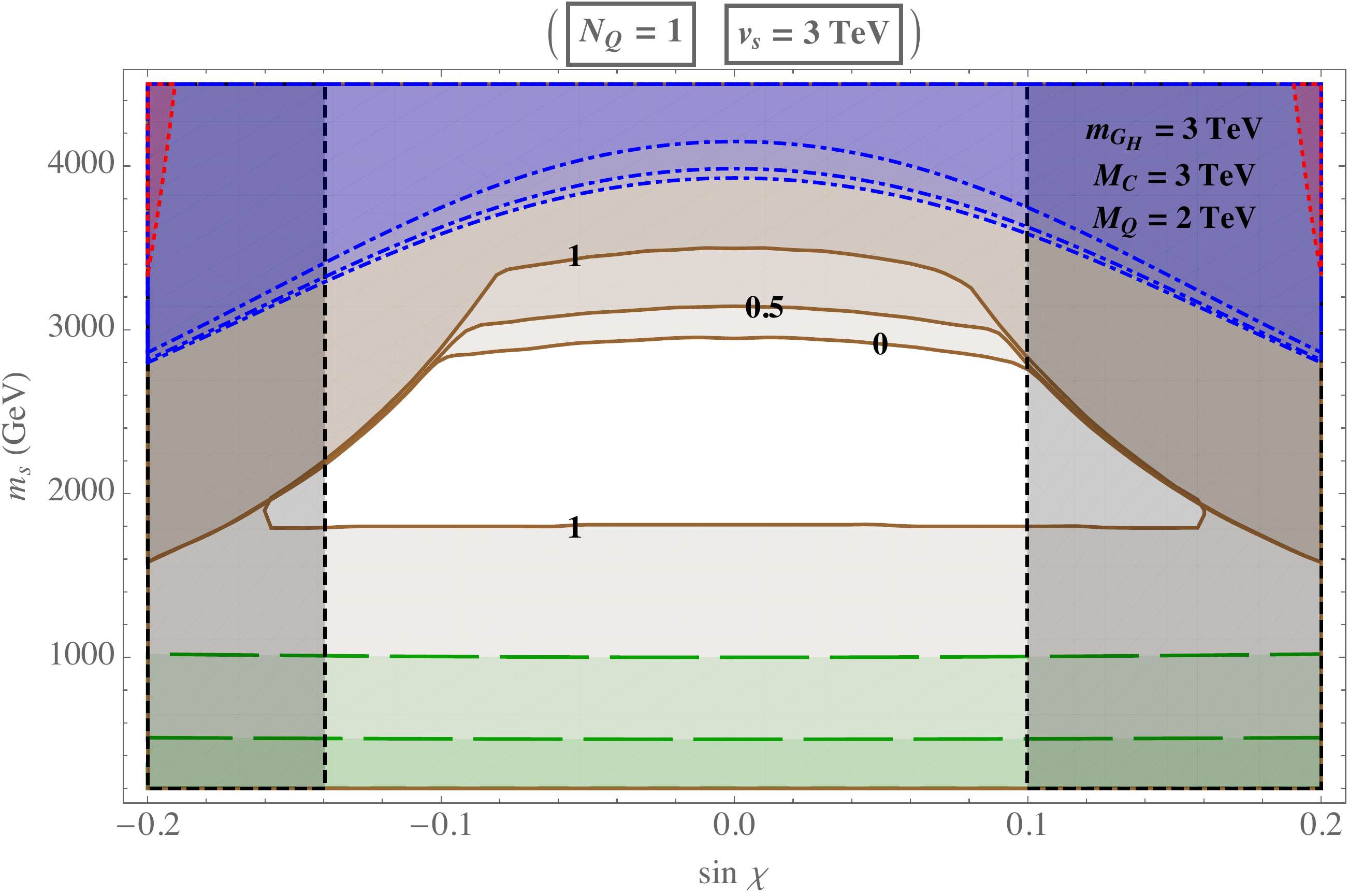}
\caption{The $\sin\chi - m_s$ exclusion plots with 1 spectator fermion generation, $N_{Q}=1$, for $v_{s}= m_{G_H} = M_C = 3$~TeV. The panels correspond to two representative values of the spectator mass, $M_Q=1,2$~TeV, and various $n \equiv m_{\mathcal A}/m_{G_H}$ curves between 0 and 1 (together with their corresponding unitarity and the global-minimum constraints from bottom to top) are displayed. Note that in the right panel an additional $m_s$~\textit{lower} bound is developed for a heavy pseudoscalar ($n=1$). All colored regions are excluded. (see the caption of Fig.~\ref{resultNQ0vs3} for further details)}
\label{resultNQ1}
\end{figure}

Examples of the viable region of the parameter space in this scenario are displayed in Fig.~\ref{resultNQ1} for $v_{s}= m_{G_H} = M_C = 3$~TeV, and selected spectator masses $M_Q =1$~and~2~TeV, as well as covering the allowed range of the pseudoscalar masses. The sources of the upper bounds on $m_s$ as a function of $\sin\chi$, including the presence of kinks in the boundary curves, are similar to what was discussed in Section \ref{NQ0}; likewise, the shape and the behavior of the allowed region as a function of the varying bosonic masses remains similar. Nevertheless, it is evident from the right panel of Fig.~\ref{resultNQ1} that a moderately heavier spectator fermion slightly reduces the largest allowed values of $m_s$. This effect is encoded in the contribution of the spectator fermions' Yukawa coupling within the scalar $\beta$-functions~\eqref{bquartic}. As was the case with the extended gauge couplings (albeit with the opposite sign), the spectator Yukawa couplings enter the scalar $\beta$-functions via two competing terms: a negative quartic term $\propto -N_{Q} \, y_{Q}^{4}$ and a positive quadratic term $\propto + N_{Q} \,y_{Q}^{2} \, \lambda_{\text{quartic}}$. For spectator masses around the singlet VEV, $M_Q \sim v_s$, the starting value of the Yukawa coupling is relatively small (c.f. \eqref{MQ}), and the positive quadratic term (multiplied by the potentially sizable scalar quartic coupling) may dominate. This enhances the positive running of the scalar quartic coupling, hastening the development of its Landau pole. As in the case of the extended gauge couplings and the coloron mass, this behavior reverses for larger spectator masses, since the negative quartic Yukawa term becomes more prominent for larger threshold values.

Interestingly, at larger values of $M_Q$ another effect is introduced, leading to a lower bound on $m_s$, as depicted in the right panel of Fig.~\ref{resultNQ1}. Recall that a heavy pseudoscalar ($n=1$) induces a small threshold value for the $\kappa_s$~coupling according to \eqref{scalarMatch}. Furthermore, for small values of the scalar quartic coupling, the positive quadratic Yukawa term in the $\beta$-function of $\kappa_s$ becomes subdominant as compared with the negative quartic Yukawa term. Hence, for (moderately) larger spectator masses together with heavy pseudoscalars, $\kappa_s$ starts small and rapidly declines in value due to the negative fermionic contributions to its running. Eventually, it becomes negative and destabilizes the potential, per \eqref{stab}. A larger $m_s$ is, therefore, necessary in order to compensate for the negative fermionic contributions in its $\beta$-function.  Above this minimum value of $m_s$, a ``window" of viable parameter space is opened, until $m_s$ becomes large enough to trigger  a Landau pole for $\lambda_s'$.   In this scenario, the upper limit on $m_s$ is still determined by the two competing effects of the triviality of $\lambda_{h}$ and $\lambda_s'$, whereas its lower bound is set by the vacuum stability condition due to the positivity of $\kappa_s$. Heavier spectators rapidly destabilize the potential via their negative fermionic Yukawa contributions to the $\kappa_s$ $\beta$-function, and close down the allowed window even when the pseudoscalars are light. In consequence, in this scenario, the spectator masses cannot exceed the singlet VEV.

\begin{figure}
\includegraphics[width=.49\textwidth]{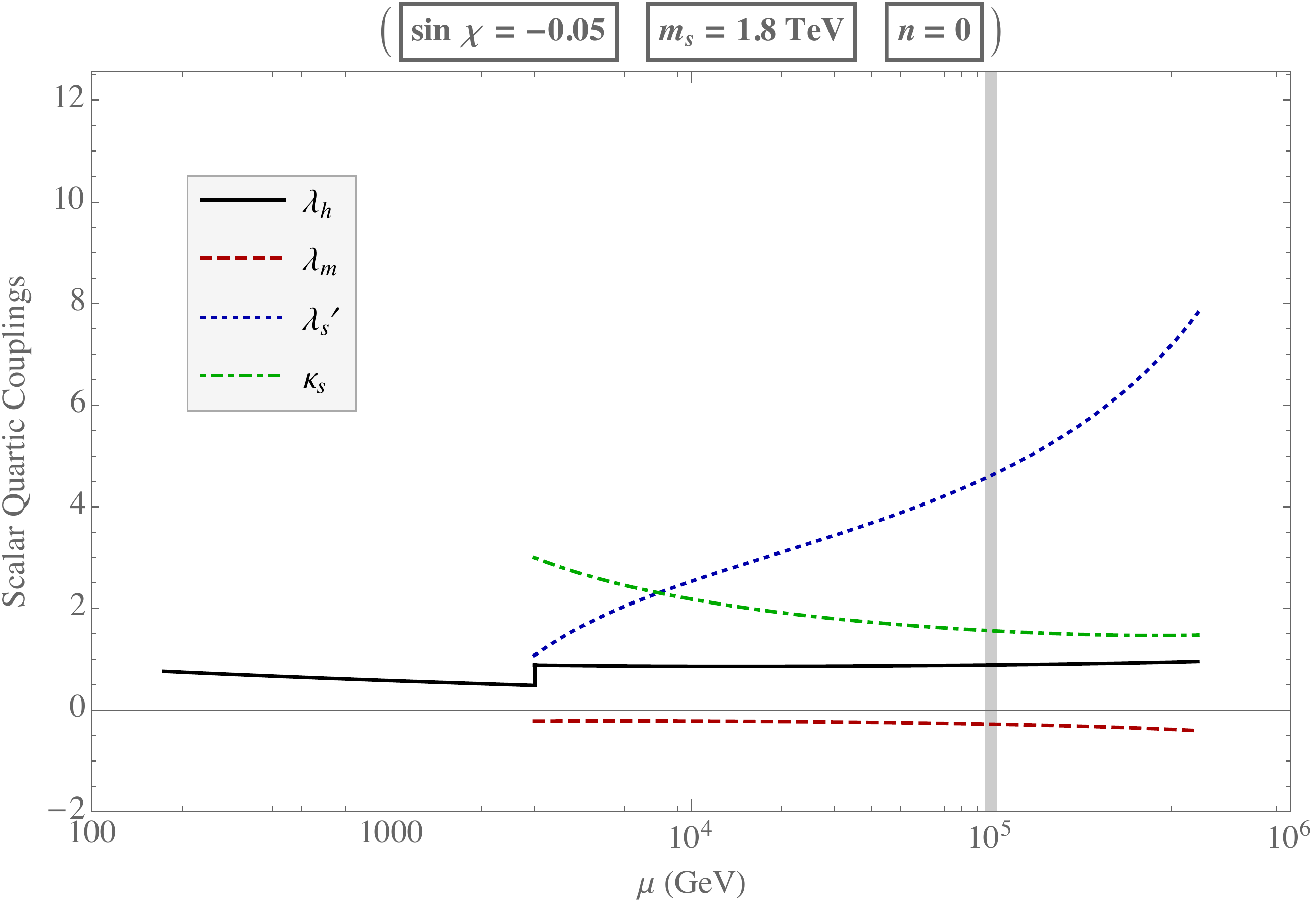}
\includegraphics[width=.49\textwidth]{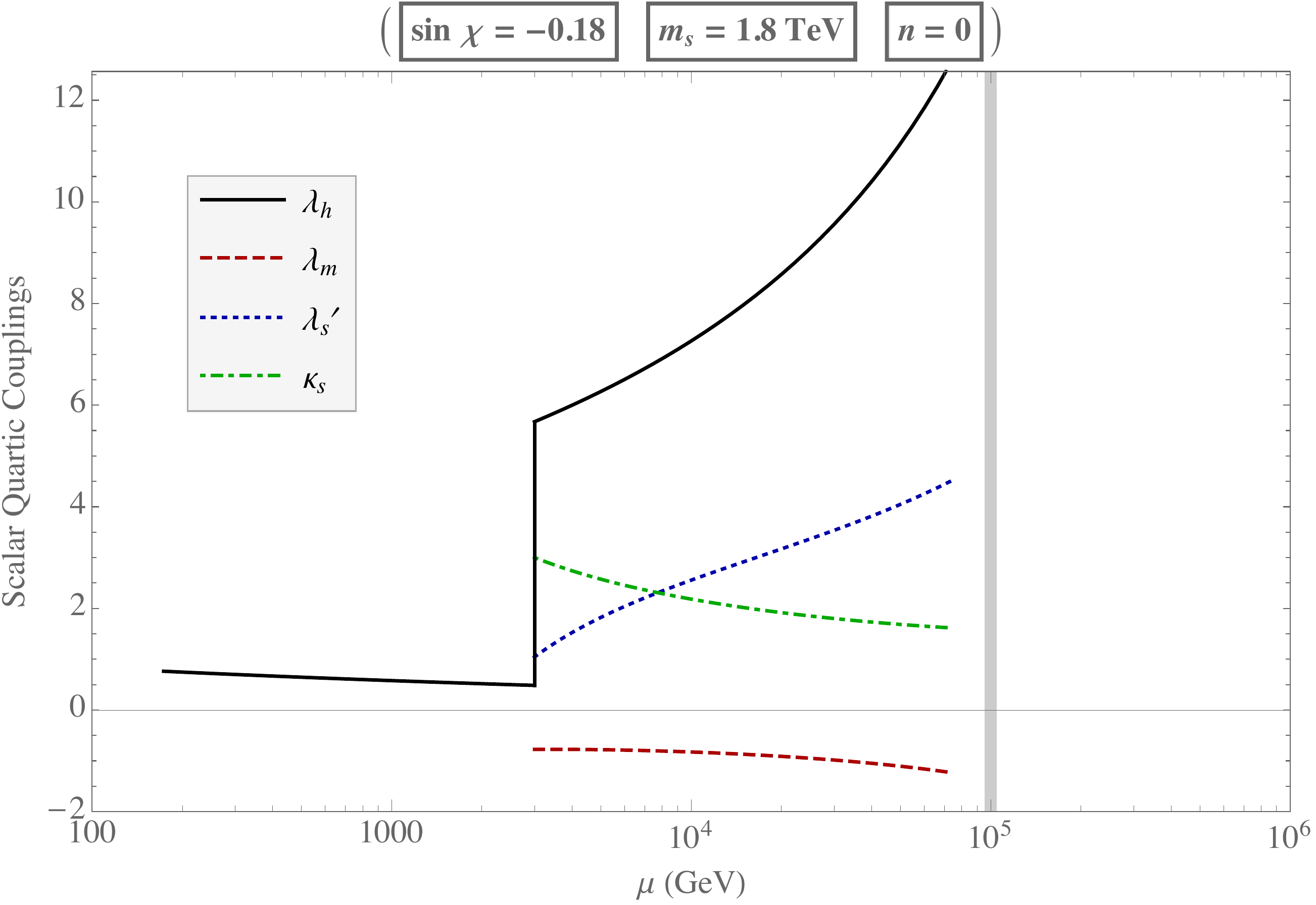}
\includegraphics[width=.49\textwidth]{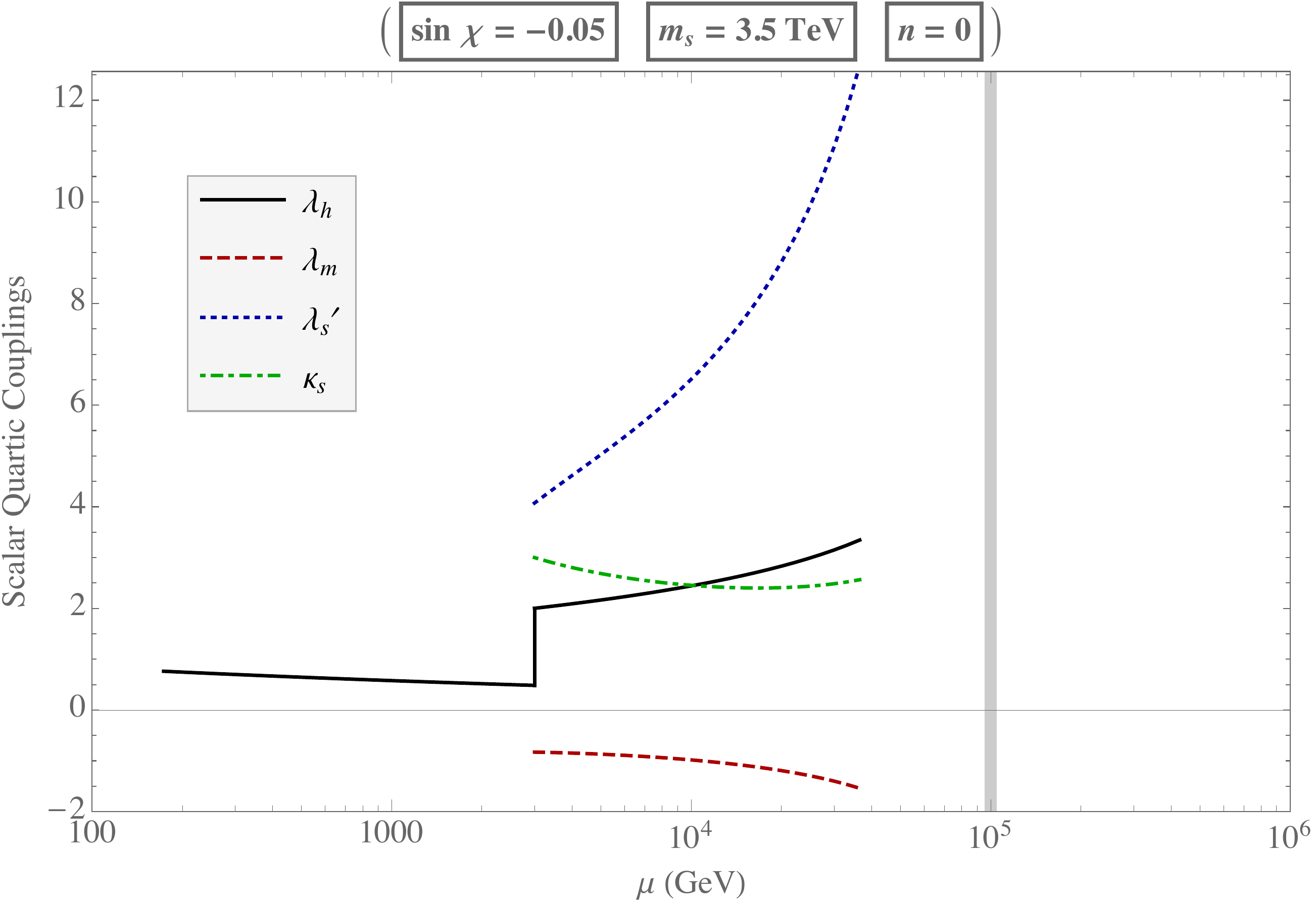}
\includegraphics[width=.49\textwidth]{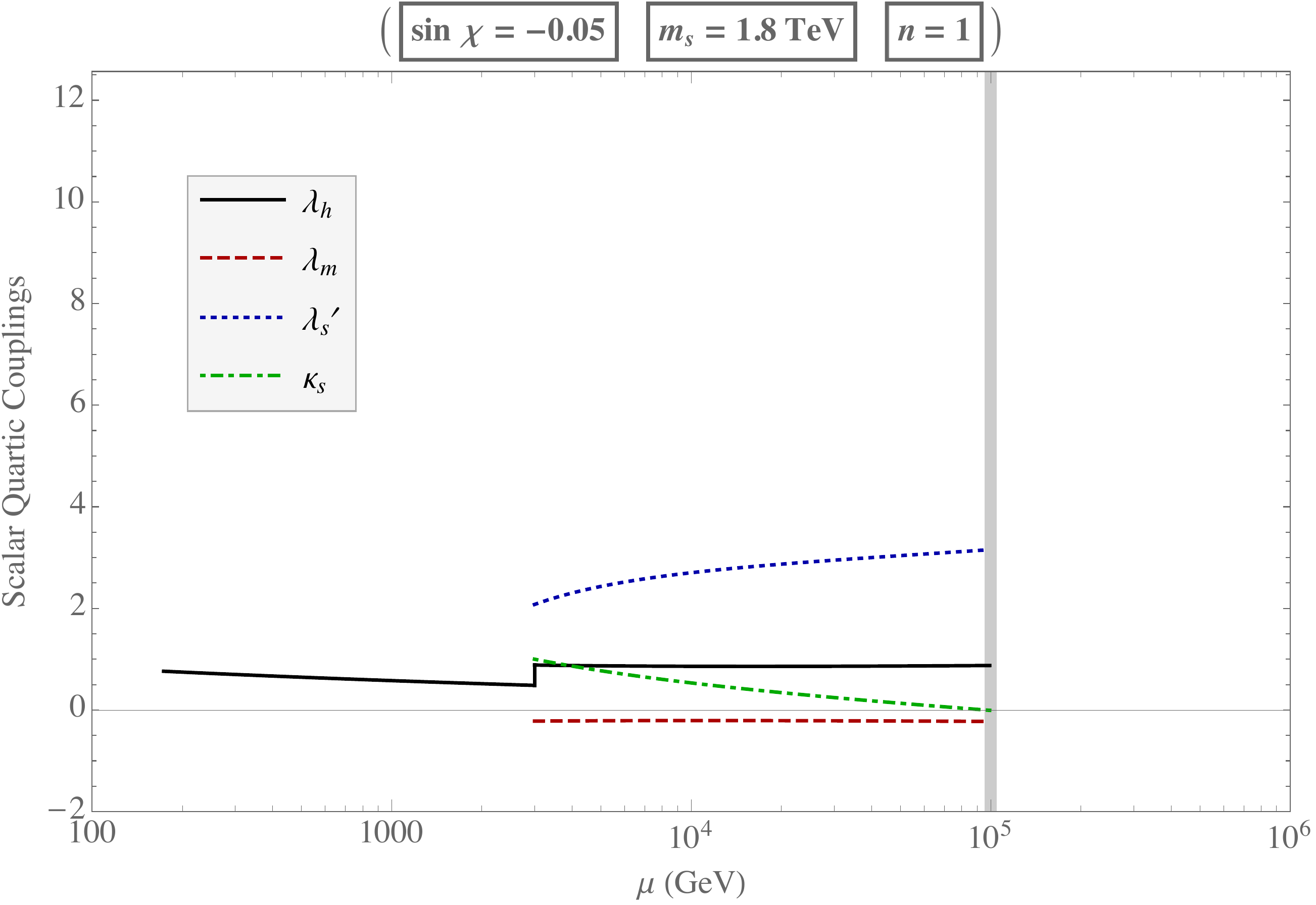}
\includegraphics[width=.5\textwidth]{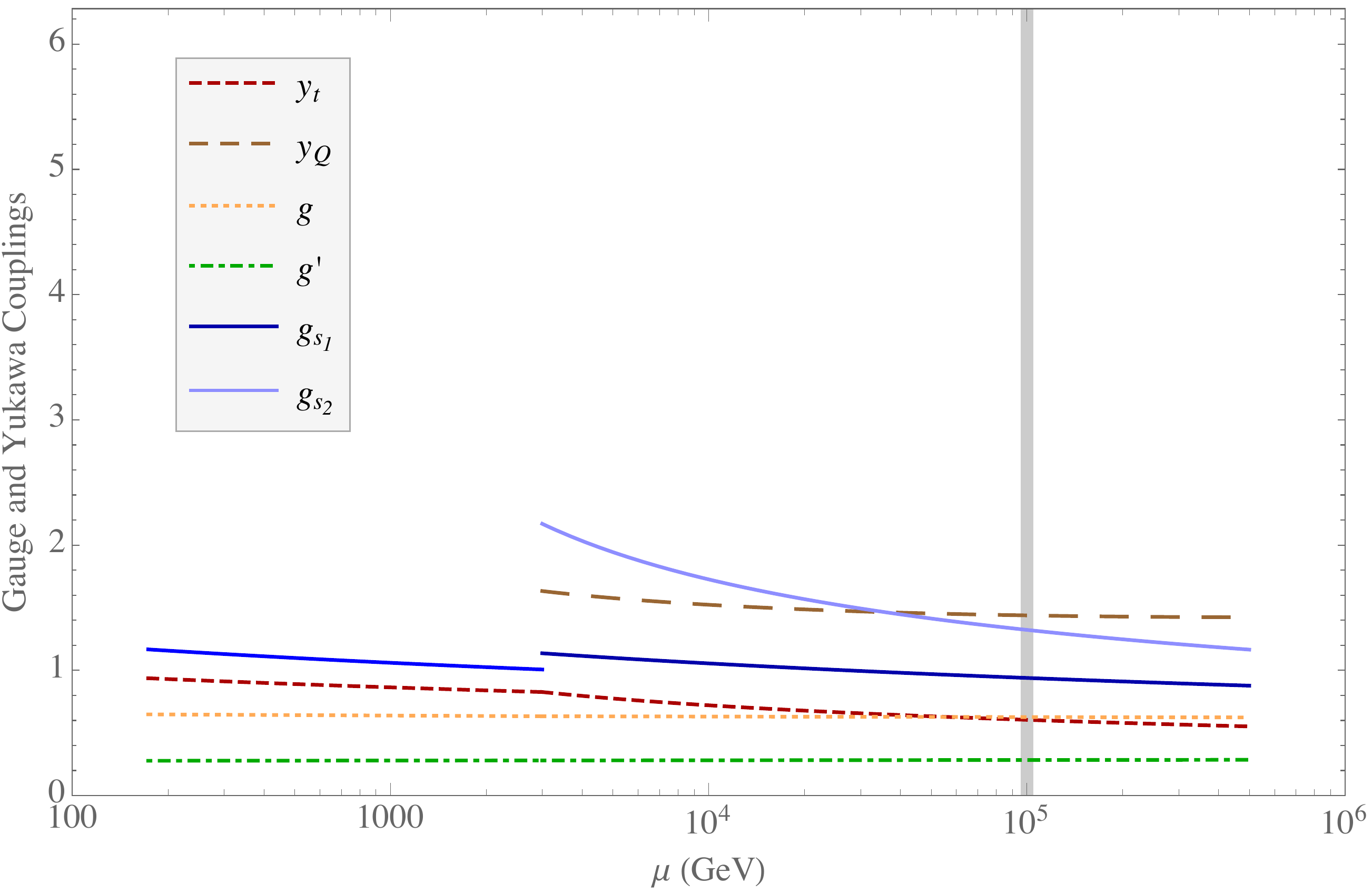}
\caption{The running of the scalar quartic couplings (top and middle rows) and the gauge and Yukawa couplings (bottom panel) as a function of the renormalization scale $\mu$, for $N_{Q}=1$, $v_{s}=m_{G_H} = M_C = 3$~TeV, and $M_Q = 2$~TeV (right panel of Fig.~\ref{resultNQ1}). The panels in the top and middle rows correspond to different regions of the $\sin\chi - m_s$ parameter space: the top-left panel illustrates a point in the allowed region for a light pseudoscalar, the top-right panel corresponds to a point excluded because $\sin\chi$ is too large ($\lambda_h$ suffers a Landau pole), and the middle-left panel corresponds to a point excluded because $m_s$ is too large ($\lambda_s'$ suffers a Landau pole). The middle-right panel demonstrates a point which is allowed for a light pseudoscalar but becomes excluded for a heavy pseudoscalar, where $\kappa_s$ becomes negative and therefore destabilizes the potential. (See the caption of Fig.~\ref{couplingsNQ0} for further details.)}
\label{couplingsNQ1}
\end{figure}

The behavior of the running couplings as a function of the renormalization scale, $\mu$, is demonstrated in Fig.~\ref{couplingsNQ1} for different parameter space regions of the right panel in Fig.~\ref{resultNQ1}. As with the previous scenario, a benchmark value within the allowed region (corresponding to a light pseudoscalar) is selected for the top-left panel, where it is shown that the vacuum stability and triviality conditions are safely satisfied up to a 100~TeV cutoff. A larger mixing angle (top-right panel) leads the Higgs quartic coupling, $\lambda_{h}$, to develop a Landau pole and excludes the corresponding parameter space. On the other hand, an overly large $m_s$ (middle-left panel) yields a Landau pole for $\lambda_{s}'$. Finally, a heavy pseudoscalar, together with the sizable spectator mass, forces the $\kappa_{s}$ coupling to become negative prematurely; the middle-right panel displays this destabilizing effect on the potential, which excludes the corresponding region of the parameter space that had been allowed for light pseudoscalars.

\subsection{Scenario with 3 Spectator Fermion Generations with Yukawa Interactions ($N_{Q}=3$)}\label{NQ3}

This scenario is described by charging all left-handed ordinary quarks under one of the $SU(3)_{i\, c}$ color groups, while the corresponding right-handed quarks are charged under the other color group. As a consequence, 3 generations of spectator fermions with opposite chiral color charges to the ordinary quarks, and Yukawa interactions with the $\Phi$~scalar, are required to cancel the anomalies. In this scenario, the $\beta$-functions of the couplings, above the threshold $v_s$, are given in Appendix~\ref{ReCoMBetaFun} by setting $N_{Q}=3$.

\begin{figure}
\includegraphics[width=.5\textwidth]{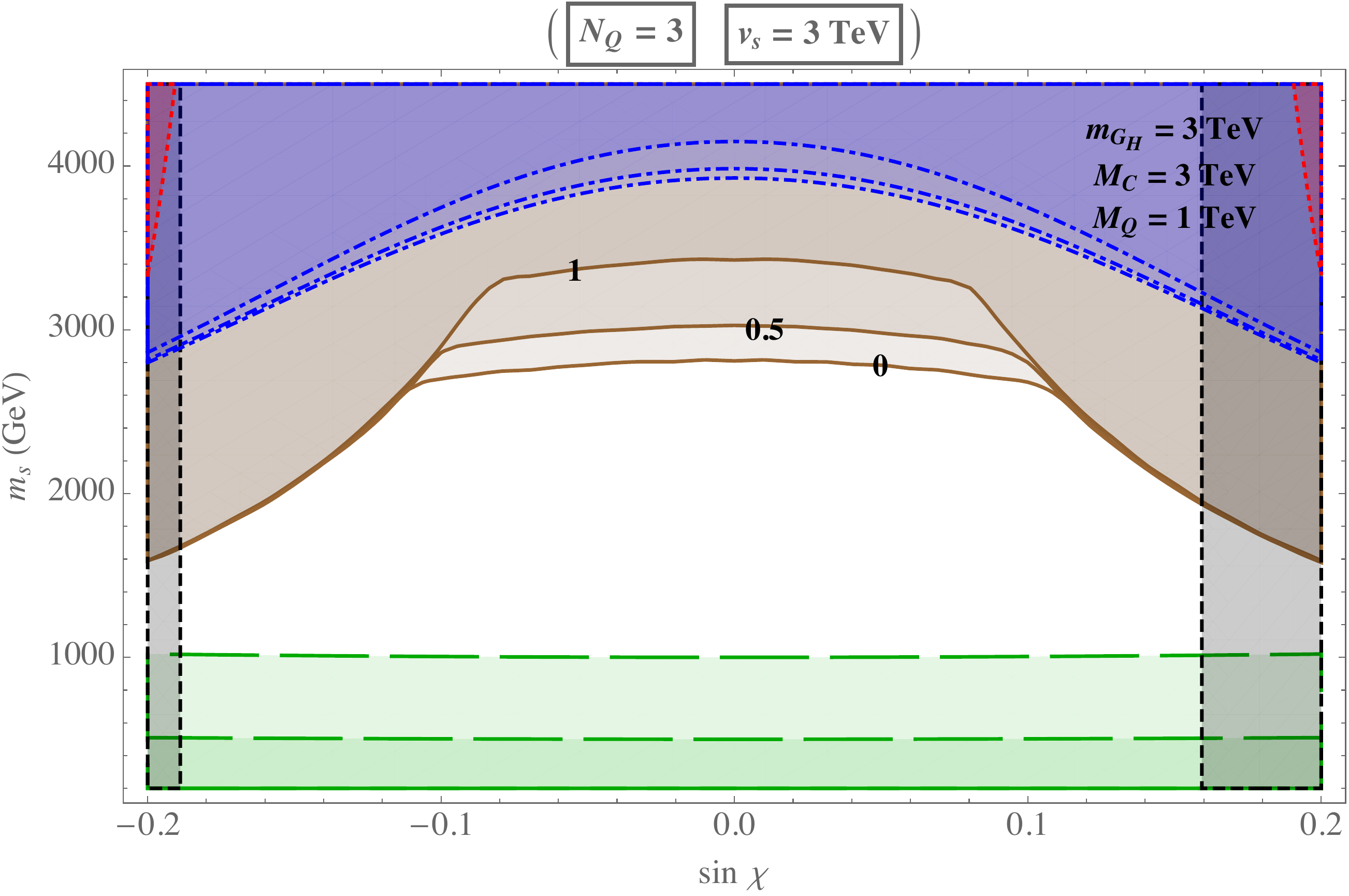}
\caption{The $\sin\chi - m_s$ exclusion plots with 3 spectator fermion generation, $N_{Q}=3$, for $v_{s}= m_{G_H} = M_C = 3$~TeV, and the spectator mass $M_Q=1$~TeV. Various $n \equiv m_{\mathcal A}/m_{G_H}$ curves between 0 and 1 (together with their corresponding unitarity and the global-minimum constraints from bottom to top) are displayed. All colored regions are excluded. (See the caption of Fig.~\ref{resultNQ0vs3} for further details.)}
\label{resultNQ3}
\end{figure}

The general observations in Section~\ref{NQ1} about the scenario containing 1 spectator fermion generation apply here as well, but are modified by the fact that the spectator Yukawa contributions to the scalar $\beta$-functions are now enhanced by the larger number of generations, $N_{Q}=3$. Fig.~\ref{resultNQ3} exhibits the viable parameter space in this scenario for the benchmarks $v_{s}= m_{G_H} = M_C = 3$~TeV, and a spectator mass $M_Q =1$~TeV. Once more, curves corresponding to the full allowed range of pseudoscalar masses are displayed. For a moderate spectator mass, the factor of 3 enhancement of its Yukawa contributions has only a moderate effect on the viable parameter space as compared with the previous scenario with only 1 spectator fermion generation. However, as the spectators become heavier, their negative fermionic contributions overwhelm the bosonic contributions within the $\beta$-function of the $\kappa_{s}$ coupling, rapidly destabilizing the potential. Hence, this scenario favors relatively light spectator fermions with masses well below $v_s$.

\subsection{Scenario with 1 Spectator Fermion Generation with Dirac Masses ($N_{Q}^{\text{Dirac}}=1$)}\label{NQ1D}

If the theory is anomaly free by construction (i.e., with the ordinary quarks vectorially charged under the extended color group, as in the scenario of Section~\ref{NQ0}), it is still possible to include spectator fermions within the renormalizable coloron model, with a (flavor universal) Dirac mass term. In this case, the Yukawa interaction \eqref{Lferm} is negligible or entirely absent. Hence, the spectator fermion mass, $M_{Q}$, is no longer related to the singlet VEV scale, $v_{s}$, and does not need to be of the same order in magnitude.  Such an example with 1 spectator generation, $N_{Q}^{\text{Dirac}}=1$, was previously studied in \cite{Top-Coloron}, where the spectator fermion generation and the third quark generation were vectorially charged under $SU(3)_{1}$ and the remaining two lighter quark generations were vectorially charged under the $SU(3)_{2}$ gauge group. In this scenario, one thus has two independent mass thresholds; namely the fermionic threshold $M_{Q}$ and the bosonic threshold characterized by $v_s$. The $\beta$-functions of this scenario (encompassing both possibilities $v_s \gtrless M_{Q}$) are provided in Appendix~\ref{DiracBetaFun}.

Fig.~\ref{resultNQ1D} exhibits the $\sin\chi - m_s$ exclusion plots for three benchmark values of the singlet VEV, $v_{s}= 3,5,10$~TeV and their corresponding vector and scalar color-octet masses, and an illustrative spectator fermion mass $M_{Q}=1$~TeV. It is interesting to observe that the panels in this figure bear a noticeable resemblance to the panels in Figs.~\ref{resultNQ0vs3}~and~\ref{resultNQ0vs510}, the scenario with no spectator fermions. This is perhaps not surprising, since the spectator Yukawa couplings are absent in both scenarios. The only difference arises in the spectator contributions to the running of the gauge couplings in the current scenario, which have a relatively small, but nonetheless noticeable, impact on the viable parameter space. Given the minor influence of the running gauge couplings on the viable parameter space, sensitivity to the exact value of the fermionic threshold, $M_{Q}$, appears insignificant.\footnote{We have checked that substituting a heavy $M_{Q}=20$~TeV does not noticeably alter the viable parameter space of Fig.~\ref{resultNQ1D}.} Moreover, the discussions in Section~\ref{NQ0}, regarding the vacuum stability and triviality analyses, still apply here. We conclude that, given the small overall effects of the running gauge couplings, inclusion of Dirac spectator fermions does not have a significant impact on the vacuum stability and triviality bounds of the renormalizable coloron model.

\begin{figure}
\includegraphics[width=.329\textwidth]{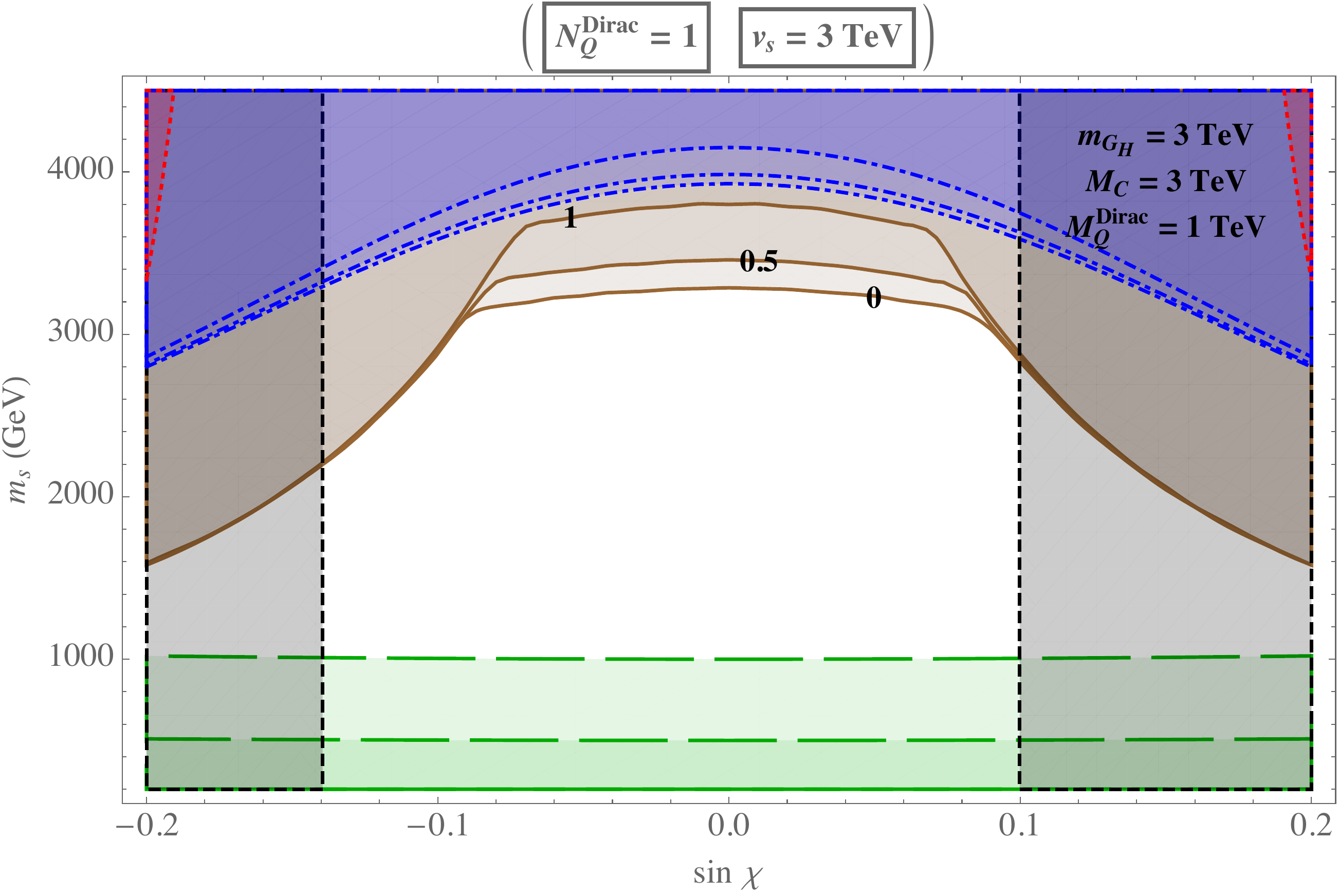}
\includegraphics[width=.329\textwidth]{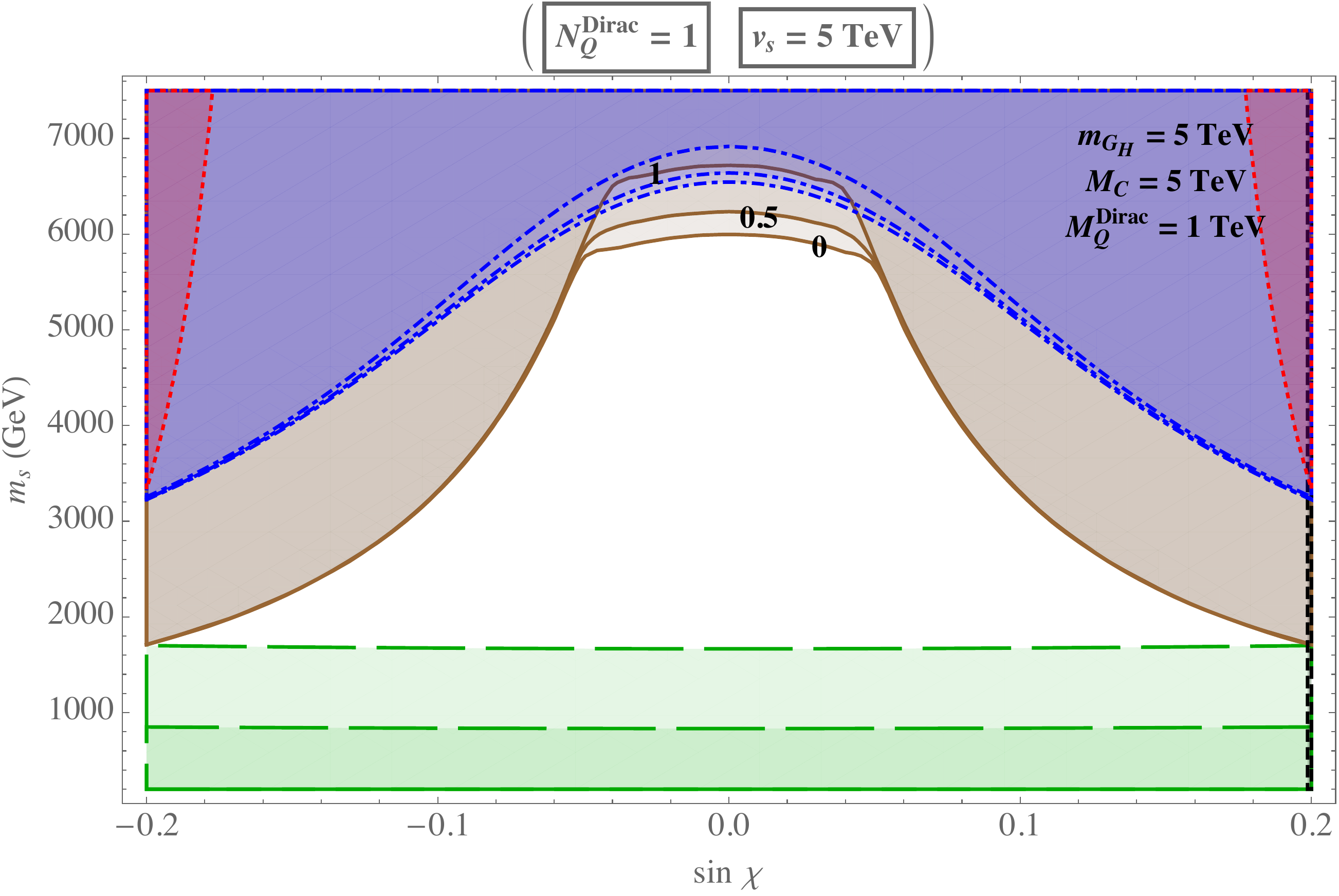}
\includegraphics[width=.329\textwidth]{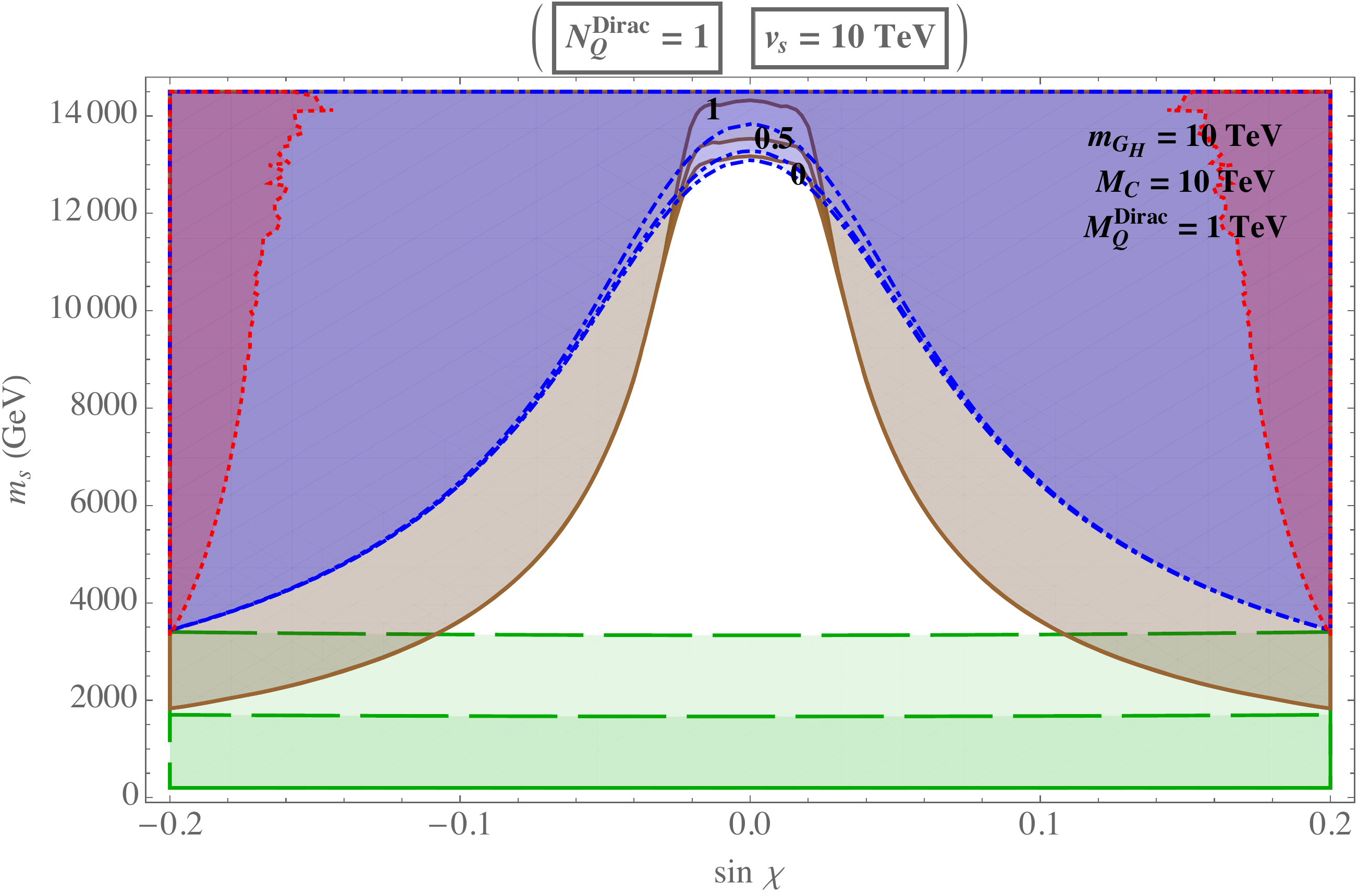}
\caption{The $\sin\chi - m_s$ exclusion plots with 1 Dirac spectator fermion generation, $N_{Q}^{\text{Dirac}}=1$, for three benchmark values $v_{s}= 3,5,10$~TeV, and the corresponding values for the vector and scalar color-octet masses. Several curves within the allowed range of the pseudoscalar masses ($n=0-1$) are presented. A universal spectator mass $M_{Q}=1$~TeV is selected for illustration. All colored regions are excluded. (See the caption of Fig.~\ref{resultNQ0vs3} for details.)}
\label{resultNQ1D}
\end{figure}

\section{Conclusion}

We have analyzed how vacuum stability and triviality requirements affect the viable parameter space of the renormalizable coloron model. To this end, we computed the $\beta$-functions of all the relevant fermionic and bosonic degrees of freedom within the formalism (presented in the Appendices), and determined the behavior of the couplings as a function of the renormalization scale.  We guaranteed the stability of the scalar potential by imposing the conditions \eqref{stab} on the running quartic couplings, while simultaneously demanding that none of the couplings encounter a Landau pole, defined as a coupling value exceeding $4\pi$. We defined the viable region of the free parameter space of the theory as the region in which these vacuum stability and triviality conditions remain satisfied up to the 100~TeV cutoff scale. Our results are summarized in exclusion plots that include complementary prior theoretical and experimental bounds from the potential's global minimum condition, unitarity, electroweak precision tests, and the LHC measurements of the 125~GeV Higgs couplings.

Our analysis covered four separate scenarios with distinctive spectator fermion contents.  The first three included 0, 1, or 3 generations of spectator fermions with Yukawa couplings to the scalar sector of the renormalizable coloron model; the fourth scenario includes a single spectator fermion generation with only Dirac masses. Our investigations revealed that the vectorial scenarios containing either no spectator fermions or 1 generation of the Dirac spectator fermions leave ample viable regions of parameter space.  The upper bounds our analysis places on the mass of the extra scalar state, as a function of the mixing between the two scalars in the model, arise because one or more scalar couplings encounters a Landau pole.  In contrast, the chiral scenarios featuring either 1 or 3 spectator generations with Yukawa couplings strongly favor relatively light spectators with masses well below the symmetry breaking scale characterized by the singlet VEV (corresponding to small Yukawa couplings).  Again, there are upper bounds on the mass of the extra scalar due to triviality.  However, if the spectator mass is increased to lie not far below the singlet VEV scale (corresponding to moderate Yukawa couplings) then the upper bounds on the scalar mass due to triviality are accompanied by lower bounds due to vacuum stability, so that the model is viable only within a window of scalar mass vs. scalar mixing angle.  A future 100~TeV hadron collider may, therefore, be considered a natural laboratory to thoroughly explore the structure of the renormalizable coloron model as a potential description of nature within this energy range. We will address an analysis of the vacuum stability and triviality up to the Planck scale elsewhere.

\section*{Acknowledgments}

The work of A.F. was supported by the IBS under the project code, IBS-R018-D1. The work of. R.S.C. and E.H.S. was supported by the National Science Foundation under Grant No. PHY-0854889. We thank Florian Lyonnet for comments on correcting a typo within the $\beta$-functions in the early version of the manuscript and Kirtimaan Mohan for assistance in verifying the $\beta$-functions in the final version of the manuscript.  R.S.C. and E.H.S. also acknowledge the support of NSF Grant \# PHYS-1066293  and the hospitality of the Aspen Center for Physics during the writing of this paper.

\appendix

\section{Quark Charges and the Top-Quark Mass
in the Renormalizable Coloron Model}
\label{sec:appx-quark-charge}

In this appendix we give a brief overview of possible fermion charges in the renormalizable coloron model, and a description of the origin of the  top-quark mass in these scenarios.

\subsection{Quark Charges}

While the charges of the quarks
 under the standard model $SU(3)_c \times SU(2)_L \times U(1)_Y$ gauge group are fixed, the color charges of the quarks can arise from either $SU(3)_{1\, c}$ or $SU(3)_{2\, c}$ in the renormalizable coloron model.
Care must be taken in the assignment of the $SU(3)_{1\, c} \times SU(3)_{2\, c}$ charges to avoid gauge anomalies, however, and some anomaly free representations include fermions with exotic color charges \cite{Frampton:1987dn}. Here we follow \cite{Cvetic:2012kv}, and investigate examples in which the only new fermions introduced are color-triplets under the unbroken $SU(3)_c$ -- and are hence color triplets under either $SU(3)_{1\, c}$ or $SU(3)_{2\, c}$.

\subsubsection{One Quark Generation}

Let us begin by discussing the case of a single Standard Model-like quark generation. By convention, we can take the left-handed weak-doublet quarks to be charged under $SU(3)_{1\, c}$, and therefore the $q_L$ state has quantum numbers $(3,1,2,+1/6)$ under the full $SU(3)_{1\, c} \times SU(3)_{2\, c} \times SU(2)_L \times U(1)_Y$ gauge group. The right-handed weak-singlet quarks, up-like ($u_R$ with hypercharge $+2/3$) and down-like ($d_R$, with hypercharge
$-1/3$), can then each be charged under either $SU(3)$ group. If both right-handed quarks are charged under $SU(3)_{1\, c}$, the representation has precisely the same form as in  the Standard Model with the replacement of $SU(3)_{1\, c}$ for $SU(3)_c$, and is automatically anomaly free.\footnote{We assume that the quark generation is accompanied by one standard model lepton generation so that all $SU(2)_L \times U(1)$ and mixed anomalies also cancel.} 

If one or both right-handed quarks transform under $SU(3)_{2\, c}$, however, the representation is anomalous and additional (spectator) fermions must be added. Following \cite{Cvetic:2012kv}, the simplest choice for the spectator fermions is to add the missing $SU(3)_{1,2\,c}$ chiral partners needed to yield an anomaly free representation. For example, if $u_R$ is charged under $SU(3)_{1\, c}$ but $d_R$ is charged under $SU(3)_{2\, c}$, we add the following down-like spectators:
\begin{equation}
d'_R:\, \, (3,1,1,-1/3)~,\ \ \ \ \ d'_L:\, \, (1,3,1,-1/3)~,
\end{equation}
{\it i.e.} add a ``$d'_R$" under $SU(3)_{1\, c}$ and a ``$d'_L$" under $SU(3)_{2\, c}$ to cancel the anomalies that were present. A similar arrangement is possible swapping $u_R \leftrightarrow d_R$, if $u_R$ is charged under $SU(3)_{2\, c}$ and $d_R$ under $SU(3)_{1\, c}$ instead.

Finally, if both $u_R$ {\it and} $d_R$ are charged under $SU(3)_{2\, c}$, there are two options. Either one can add weak-singlet $u'$ and $d'$ quarks (combining the two possiblities described above), or alternatively one can add weak-doublet spectator quarks
\begin{equation}
q'_R:\, \, (3,1,2,+1/6)~,\ \ \ \ \ q'_L:\, \, (1,3,2,+1/6)~.
\end{equation}

Note that, in all of the cases above, a Yukawa coupling to the color bifundamental $\Phi:\, (3, \bar{3},1,1)$ scalar can give mass to the additional spectator fermions that were added.\footnote{This is in contrast with many of the exotic-color scenarios described in \cite{Frampton:1987dn}, which require weak-doublet bifundamental scalars as well.}  For simplicity, in what follows and in the body of this paper we consider only the last case -- the addition of weak-doublet spectators that transform vectorially under $SU(2)_L \times U(1)_Y$. None of the properties of the renormalizable coloron model discussed in this paper depend sensitively on this choice.

\subsubsection{Three Quark Generations}

In a model, like the renormalizable coloron model, that includes three quark generations (and weak-doublet spectators), there are several distinct color charge assignments that are possible. First, one could choose \cite{Chivukula:1996yr} all left- and right-handed quarks to be charged under $SU(3)_{1\, c}$. In this case no spectator fermions are necessary. Alternatively, one could choose \cite{Top-Coloron} one left-handed weak-doublet and a single right-handed weak-singlet up- and down-quark (loosely speaking, ``one generation of quarks") to be charged under $SU(3)_{1\, c}$, and the other quarks (the ``other two generations") to be charged under $SU(3)_{2\, c}$ -- and again, no spectators are required to cancel anomalies.  While neither case involves spectator fermions, it is the first case that we have studied in the $N_Q = 0$ scenario of Sec.~\ref{NQ0}.

Second, one can have two left-handed weak doublet quarks and one right-handed weak singlet up- and down-quark be charged under $SU(3)_{1\, c}$ and the complement (one left-handed weak doublet quark, and two right-handed weak-singlet up- and down-quarks) charged under $SU(3)_{2\, c}$. In this case, one needs one ``generation" of weak-doublet spectator quarks
\begin{equation}
\Psi'_R:\, \, (3,1,2,+1/6),\ \ \ \ \ \Psi'_L:\, \, (1,3,2,+1/6)~,
\label{eq:spectator-charges}
\end{equation}
to cancel anomalies -- this is the case labeled $N_Q=1$ in the body of the paper and studied in Sec.~\ref{NQ1}.

Third, one could choose all left-handed weak doublet quarks to be charged under $SU(3)_{1\, c}$ and all right-handed up- and down-quarks to be charged under $SU(3)_{2\, c}$. In this case three generations of spectators with the quantum numbers of   (\ref{eq:spectator-charges}) are required. This case is labeled $N_Q=3$ in the body of the paper; it is explored in Sec.~\ref{NQ3}.

Finally, we note that it is always possible to add additional spectator quarks that are vectorially charged under both the color- and weak gauge groups.
One such scenario, motivated by the need to accomodate quark mixing
\cite{Top-Coloron} in a case where no spectators are required to cancel anomalies, is considered in the text in Sec.~\ref{NQ1D} and denoted $N_Q^{\rm Dirac} = 1$.

\subsection{The Top-Quark Mass}

In addition to specifying the gauge charges of the fermions, a complete analysis of the model requires determining the fermion mass eigenstates and examining the weak- and flavor-phenomenology that results from the extended gauge and scalar interactions. In general, phenomenological consistency will require that the additional scalar, fermion, and vector states be sufficiently heavy to be consistent with existing experimental data -- and, for example, not give rise to overly large flavor-changing neutral current interactions. A thorough phenomenological analysis of all the possiblities described above is beyond the scope of the current work.\footnote{For a recent weak- and flavor-phenomeology analysis in one case see \cite{Top-Coloron}.}

This paper focuses, instead, on the constraints on the renormalizable coloron model arising from high-energy vacuum stability and triviality. For the purposes of our analysis here, then, we can neglect the masses of all of the light quarks (and leptons) since their corresponding couplings are small. The top-quark Yukawa coupling is sizable, however. We outline the form of the top-quark mass generating sector of the renormalizable coloron model corresponding to the gauge charge choices described above. 

We assume here that the mass-eigenstate left- and right-handed top-quark fields are as ``aligned" with the $SU(3)_{1\, c} \times SU(3)_{2\, c}$ gauge charges as possible.
In the cases where $N_Q=0,1$, therefore, we consider the situation in which the left- and right-handed top-quark are almost entirely left- and right-handed gauge-eigenstate fields which transform under (by convention) $SU(3)_{1\, c}$. Then, the top-quark mass arises from a standard-model like Yukawa coupling \cite{Chivukula:2013xka} to the weak-doublet Higgs boson $\phi$ -- and hence the corresponding coupling constant is the same in both the low-energy ($\mu < v_s$) and high-energy 
($\mu \ge  v_s$) theories. The running of this coupling, however, differs in these two energy regimes; it has the SM form of  ~(\ref{SMBeta}) in the low-energy regime and the modified form corresponding to  ~(\ref{bYukawa}) in the high-energy regime.

The situation when $N_Q=3$ is different. In this case, the 
gauge eigenstates $t_{L,R}$ (which are charged, respectively, under
$SU(3)_{1c,2c}$) mix substantially with a set of weak-vector
spectators which we will denote $T_{L,R}$ (and are charged, respectively, under $SU(3)_{2c,1c}$). After the extended color and electroweak symmetries break, the correponding mass-mixing matrix
is of the form
\begin{equation}
{\cal L}_m = -
\begin{pmatrix}
\bar{t}_R & \bar{T}_R
\end{pmatrix}
\begin{pmatrix}
0 & \tilde{m_t}\\
m & M
\end{pmatrix}
\begin{pmatrix}
t_L \\
T_L
\end{pmatrix}~, \qquad \tilde{m_t} \equiv \frac{\tilde{y}_t v_h}{\sqrt{2}},\quad M \equiv \frac{y_Q v_s}{\sqrt{6}}~.
\end{equation}
Here, $m$ arises from a Dirac mass term coupling the left-handed
third-generation quark field to the corresponding field in $\Psi'_R$ of  ~(\ref{eq:spectator-charges}), 
\begin{equation}
m
\begin{pmatrix}
\bar{t}_L & \bar{b}_L
\end{pmatrix}
\Psi'_R~,
\end{equation}
$\tilde{y}_t$ is a high-energy Yukawa coupling of $t_R$ to $T_L$ (both are charged under $SU(3)_{2\, c}$), 
\begin{equation}
\tilde{y}_t \bar{t}_R \tilde{\phi} \Psi'_L~,
\end{equation}
and $y_Q$ is the spectator mass Yukawa coupling of  ~(\ref{Lferm}). 

In the limit $\tilde{m}_t \ll m, M$, this matrix is of seesaw form \cite{Dobrescu:1997nm,Chivukula:1998wd},
and yields a top-quark mass
\begin{equation}
m_t = \frac{\tilde{m}_t m}{M}~,
\label{eq:top-mass}
\end{equation}
with approximate mass-eigenstate fields
\begin{align}
t^{phys}_L & \approx \frac{M t_L - m T_L}{\sqrt{m^2 + M^2}}\\
t^{phys}_R & \approx t_R + {\cal O}\left(\frac{\tilde{m}_t}{M}\right) T_R~. 
\end{align}
From  ~(\ref{eq:top-mass}) we see that the low-energy
top-quark Yukawa coupling ($y_t$) is related to the high-energy
coupling through
\begin{equation}
y_t = \tilde{y}_t \frac{m}{M}~.
\end{equation}
In this limit, the corresponding spectator mass is
$\sqrt{m^2 + M^2}$ with mass eignestate fields orthogonal to those of the top-quark.

For the purposes of illustration, we consider
$m/M \simeq 1$ and impose the boundary condition $\tilde{y}_t(v_s) = y_t(v_s)$ when integrating the renormalization group equations in Sec.~\ref{NQ3} and displaying the results in Fig.~\ref{resultNQ3}.  Since, as  ~(\ref{bquartic}) reveals, the impact of the coupling $\lambda_m$ exceeds that of $y_t$ in the running of $\lambda_h$, and our results should be relatively insensitive to the value of $m/M$.

\section{One-Loop $\beta$-Functions of the SM}\label{SMBetaFun}

At low energies below the threshold scale, $v_s$, all the heavy non-SM degrees of freedom may be integrated out, and one recovers the ordinary SM as the low energy effective theory. Analytical expressions for the $\beta$-functions ($\beta_{\mathcal C}=\mu \,d\mathcal{C}/d \mu$, with $\mathcal C$ the running coupling as a function of the energy $\mu$) of the SM couplings are known up to two loops in the literature (see e.g. \cite{Schrempp:1996fb}). Here, we review the relevant one-loop SM $\beta$-functions for completeness, according to the normalization of the scalar potential \eqref{pot}:
\begin{equation} \label{SMBeta}
\begin{split}
\pbrac{4\pi}^{2}\beta_{g}^{\text{SM}} =&-g^{3} \tbrac{+\frac{19}{6}} \ , \quad \pbrac{4\pi}^{2}\beta_{g'}^{\text{SM}} =+g'^{\,3} \tbrac{+\frac{41}{6}} \ , \quad \pbrac{4\pi}^{2}\beta_{g_{s}}^{\text{SM}} = -g_{s}^{3} \tbrac{+7} \ ,\\
\pbrac{4\pi}^{2}\beta_{y_{t}}^{\text{SM}} =&\, y_{t} \tbrac{-8 g^{2}_{s} - \frac{9}{4} g^{2} - \frac{17}{12} g'^{\,2} +\frac{9}{2} y_{t}^{2}}\ , \\
\pbrac{4\pi}^{2}\beta_{\lambda_{h}}^{\text{SM}} =&+4\lambda_{h}^{2} +3\lambda_{h} \tbrac{4 y_{t}^{2} - 3g^{2} - g'^{2}}- \frac{9}{4} \tbrac{16 y_{t}^{4} -2 g^{4} - (g^{2} + g'^{2})^{2}} \ .
\end{split}
\end{equation}
We note that the hypercharge coupling is normalized according to $g' = \sqrt{3/5} \, g_{1}$, where $g_{1}$ is the corresponding coupling with the GUT normalization, and the contributions of all the light fermions, except for the top quark, are ignored.

\section{One-Loop $\beta$-Functions of the Renormalizable Coloron Model: Spectators Fermions with Yukawa Interactions}\label{ReCoMBetaFun}

In this section, we provide the general expressions for the one-loop $\beta$-functions of the gauge, Yukawa, and scalar quartic couplings, within the context of the renormalizable coloron model at energy scales above the threshold $v_s$, which characterizes the appropriate symmetry breaking and mass generating scale. The $\beta$-functions may be calculated either directly using the Feynman rules of the theory,\footnote{See e.g. the appendices in \cite{Chivukula:2013xka,Chivukula:2011ng,Chivukula:2013xla}.} or by employing the well-known general expressions in the literature \cite{GenBeta1,GenBeta2}. All $\beta$-functions are computed within the $\overline{\text{MS}}$~renormalization scheme.

At energy scales above the threshold, $v_s$, the heavy states of the renormalizable coloron model fully contribute to the running of the couplings, since they all obtain their masses due to the spontaneous symmetry breaking. In particular, this is true for the spectator fermions with a Yukawa interaction with the $\Phi$~scalar, as in \eqref{Lferm}. Moreover, above the singlet VEV scale, the extended $SU(3)_{1}\times SU(3)_{2}$ color gauge group is restored, and the low energy QCD coupling $g_{s}$ is traded for the two couplings $g_{s_{1}}$ and $g_{s_{2}}$ of the extended gauge group.

In this energy regime, the $\beta$-functions of the gauge couplings $g$, $g'$, $g_{s_{1}}$ and $g_{s_{2}}$ are given by:
\begin{equation}\label{bgauge}
\begin{split}
\pbrac{4\pi}^{2}\beta_{g} =& -g^{3} \tbrac{+\frac{19}{6}  - 2 N_{Q}}\ , \quad \pbrac{4\pi}^{2}\beta_{g'} = +g'^{\,3} \tbrac{+\frac{41}{6} +\frac{2}{9} N_{Q}} \ , \\
\pbrac{4\pi}^{2}\beta_{g_{s_{1}}} =& -g_{s_{1}}^{3} \tbrac{\begin{cases}
9&(\text{for}\, N_{Q}\neq 0)\\
7&(\text{for}\, N_{Q}=0)
\end{cases} \, - \frac{4}{3} \frac{N_{Q}}{2} -\frac{1}{2} }\ , \quad \pbrac{4\pi}^{2}\beta_{g_{s_{2}}} =-g_{s_{2}}^{3} \tbrac{\begin{cases}
9&(\text{for}\, N_{Q}\neq 0)\\
11&(\text{for}\, N_{Q}=0)
\end{cases} \, - \frac{4}{3} \frac{N_{Q}}{2} -\frac{1}{2} } \ ,
\end{split}
\end{equation}
with $N_{Q}$ denoting the number of spectator fermion generations that act chirally under the extended color gauge group. As mentioned in Section~\ref{NQ0}, the scenario with no spectator fermions ($N_{Q}=0$) is represented by charging all the ordinary quarks vectorially under the $SU(3)_{1\, c}$ color group; hence, the $SU(3)_{2\, c}$ color group does not have fermionic content in this particular scenario. The complex bi-fundamental $\Phi$~scalar, being a $(3,\bar 3)$ under the extended color gauge group, contributes equally to the running of both gauge couplings with a factor of -1/2. 

Next, we present the $\beta$-functions of the relevant Yukawa couplings in the high energy regime; namely, those of the top quark and the spectator fermion:
\begin{equation}\label{bYukawa}
\begin{split}
\pbrac{4\pi}^{2}\beta_{y_{t}} =&\, y_{t} \tbrac{ \begin{cases}
-4\pbrac{g^{2}_{s_{1}}+g^{2}_{s_{2}}}&(\text{for}\, N_{Q}\neq 0)\\
-8 g^{2}_{s_{1}}&(\text{for}\, N_{Q}=0)
\end{cases} \, - \frac{9}{4} g^{2} - \frac{17}{12} g'^{\,2} +\frac{9}{2} y_{t}^{2} }\ , \\
\pbrac{4\pi}^{2}\beta_{y_{Q}} =&\, y_{Q} \tbrac{ - 4 \pbrac{g^{2}_{s_{1}} + g^{2}_{s_{2}}} - \frac{9}{2} g^{2} -\frac{1}{6} g'^{\, 2} + \pbrac{3+2N_{Q}} y_{Q}^{2} } \ .
\end{split}
\end{equation}
Once more, we note that a mixing between the SM quarks and the spectator fermions is negligible, due to the constraints from the flavor-changing coloron couplings \cite{Top-Coloron}.

Finally, we report the expressions for the $\beta$-functions of the scalar quartic couplings above the singlet VEV scale:
\begin{equation}\label{bquartic}
\begin{split}
\pbrac{4\pi}^{2}\beta_{\lambda_{h}} =&\, +4\lambda_{h}^{2} + 54 \lambda_{m}^{2} +3\lambda_{h} \tbrac{4 y_{t}^{2} - 3g^{2} - g'^{2}}- \frac{9}{4} \tbrac{16 y_{t}^{4} - 2 g^{4} - (g^{2} + g'^{2})^{2}}  \ , \\
\pbrac{4\pi}^{2}\beta_{\lambda_{m}} =&\, \lambda_{m} \tbrac{ +4\lambda_{m} + 2\lambda_{h} +\frac{20}{3}\lambda_{s}' + \frac{16}{3} \kappa_{s} + \frac{3}{2} \tbrac{4 y_{t}^{2} - 3g^{2} - g'^{2}} + 4\tbrac{ N_{Q} \,y_{Q}^{2} - 2 \pbrac{g^{2}_{s_{1}} + g^{2}_{s_{2}}}}} \ , \\
\pbrac{4\pi}^{2}\beta_{\lambda_{s}'} =&\, +\frac{26}{3}\lambda_{s}'^{\, 2} +12\lambda_{m}^{2} +\frac{32}{3}\kappa_{s}^{2} +\frac{32}{3}\lambda_{s}' \kappa_{s} + 8\lambda_{s}' \tbrac{ N_{Q} \,y_{Q}^{2} - 2 \pbrac{g^{2}_{s_{1}} + g^{2}_{s_{2}}}} - 8\tbrac{ N_{Q} \, y_{Q}^{4} -\pbrac{g^{2}_{s_{1}} + g^{2}_{s_{2}}}^{2}} \ , \\
\pbrac{4\pi}^{2} \beta_{\kappa_{s}} =&\,  +8\kappa_{s}^{2} + 4 \kappa_{s} \lambda_{s}' + 8\kappa_{s} \tbrac{ N_{Q} \,y_{Q}^{2} - 2 \pbrac{g^{2}_{s_{1}} + g^{2}_{s_{2}}}} - 4\tbrac{ 2N_{Q} \, y_{Q}^{4} -\frac{5}{8}\pbrac{g^{4}_{s_{1}} + g^{4}_{s_{2}}} + g^{2}_{s_{1}} g^{2}_{s_{2}}} \ .
\end{split}
\end{equation}

\section{One-Loop $\beta$-Functions of the Renormalizable Coloron Model: Spectators Fermions with Dirac Masses}\label{DiracBetaFun}

The details of this scenario are described in Section~\ref{NQ1D}. In the energy regime lower than the bosonic and fermionic mass scales, one recovers the SM effective theory and its $\beta$-functions \eqref{SMBeta}. Here, we provide the $\beta$-functions pertaining to this scenario at higher energies, giving separate attention to cases where the spectator fermion may be heavier or lighter than the bosonic states.

\subsection{Spectator Fermions Lighter than the Bosonic States ($M_{Q} <\mu < v_s$)}

In the energy regime between the fermion and boson masses, $M_{Q} <\mu < v_s$, the theory effectively corresponds to the SM augmented by 1 generation of left- and right-handed spectator fermions, which directly influences the gauge coupling $\beta$-functions:
\begin{equation} \label{SMandFerm}
\pbrac{4\pi}^{2}\beta_{g} =-g^{3} \tbrac{+\frac{19}{6}-2} \ , \quad \pbrac{4\pi}^{2}\beta_{g'} =+g'^{\,3} \tbrac{+\frac{41}{6} + \frac{2}{9}} \ , \quad \pbrac{4\pi}^{2}\beta_{g_{s}} = -g_{s}^{3} \tbrac{+7-\frac{4}{3}} \ .
\end{equation}
The top quark and the $\lambda_{h}$ quartic coupling $\beta$-functions remain unaffected and retain their SM forms \eqref{SMBeta}.

\subsection{Spectator Fermions Heavier than the Bosonic States ($v_s <\mu < M_{Q}$)}

In the energy regime between the fermion and boson masses, $v_s <\mu < M_{Q}$, the theory corresponds to the SM augmented by the extended color gauge group and the extended scalar sector, sans the spectator fermions. With the third quark generation vectorially charged under $SU(3)_{1}$ and the remaining two lighter quark generations vectorially charged under $SU(3)_{2}$, the $\beta$-functions take the following form:
\begin{equation}\label{SMandBos}
\begin{split}
\pbrac{4\pi}^{2}\beta_{g} =& -g^{3} \tbrac{+\frac{19}{6}}\ , \quad \pbrac{4\pi}^{2}\beta_{g'} = +g'^{\,3} \tbrac{+\frac{41}{6}} \ , \\
\pbrac{4\pi}^{2}\beta_{g_{s_{1}}} =& -g_{s_{1}}^{3} \tbrac{11-\frac{4}{3} -\frac{1}{2} }\ , \quad \pbrac{4\pi}^{2}\beta_{g_{s_{2}}} =-g_{s_{2}}^{3} \tbrac{11 - \frac{8}{3} -\frac{1}{2} } \ , \\
\pbrac{4\pi}^{2}\beta_{y_{t}} =&\, y_{t} \tbrac{-8 g^{2}_{s_{1}} - \frac{9}{4} g^{2} - \frac{17}{12} g'^{\,2} +\frac{9}{2} y_{t}^{2}}\ , \\
\pbrac{4\pi}^{2}\beta_{\lambda_{h}} =&\, +4\lambda_{h}^{2} + 54 \lambda_{m}^{2} +3\lambda_{h} \tbrac{4 y_{t}^{2} - 3g^{2} - g'^{2}}- \frac{9}{4} \tbrac{16 y_{t}^{4} - 2 g^{4} - (g^{2} + g'^{2})^{2}}  \ , \\
\pbrac{4\pi}^{2}\beta_{\lambda_{m}} =&\, \lambda_{m} \tbrac{ +4\lambda_{m} + 2\lambda_{h} +\frac{20}{3}\lambda_{s}' + \frac{16}{3} \kappa_{s} + \frac{3}{2} \tbrac{4 y_{t}^{2} - 3g^{2} - g'^{2}}  - 8 \pbrac{g^{2}_{s_{1}} + g^{2}_{s_{2}}}} \ , \\
\pbrac{4\pi}^{2}\beta_{\lambda_{s}'} =&\, +\frac{26}{3}\lambda_{s}'^{\, 2} +12\lambda_{m}^{2} +\frac{32}{3}\kappa_{s}^{2} +\frac{32}{3}\lambda_{s}' \kappa_{s} -16 \lambda_{s}'  \pbrac{g^{2}_{s_{1}} + g^{2}_{s_{2}}} + 8\pbrac{g^{2}_{s_{1}} + g^{2}_{s_{2}}}^{2} \ , \\
\pbrac{4\pi}^{2} \beta_{\kappa_{s}} =&\,  +8\kappa_{s}^{2} + 4 \kappa_{s} \lambda_{s}' -16\kappa_{s}  \pbrac{g^{2}_{s_{1}} + g^{2}_{s_{2}}} +\frac{5}{2}\pbrac{g^{4}_{s_{1}} + g^{4}_{s_{2}}} -4 g^{2}_{s_{1}} g^{2}_{s_{2}} \ .
\end{split}
\end{equation}

\subsection{Above the Fermionic and Bosonic Mass Scales ($\mu > v_s, M_{Q}$)}

In the energy regime above the fermion and boson masses, $\mu > v_s, M_{Q}$, all heavy states contribute to the running of the couplings. The gauge coupling $\beta$-functions are of this form:
\begin{equation}\label{SMandEverything}
\begin{split}
\pbrac{4\pi}^{2}\beta_{g} =& -g^{3} \tbrac{+\frac{19}{6}-2}\ , \quad \pbrac{4\pi}^{2}\beta_{g'} = +g'^{\,3} \tbrac{+\frac{41}{6}+\frac{2}{9}} \ , \\
\pbrac{4\pi}^{2}\beta_{g_{s_{1}}} =& -g_{s_{1}}^{3} \tbrac{11-\frac{4}{3}-\frac{4}{3} -\frac{1}{2} }\ , \quad \pbrac{4\pi}^{2}\beta_{g_{s_{2}}} =-g_{s_{2}}^{3} \tbrac{11 - \frac{8}{3} -\frac{1}{2} } \ ,
\end{split}
\end{equation}
whereas, the top quark and the scalar quartic coupling $\beta$-functions have the same form as in \eqref{SMandBos}.

\end{document}